\documentclass[aps,prd,twocolumn,superscriptaddress,nofootinbib,longbibliography]{revtex4-2}
\usepackage[T1]{fontenc}
\usepackage{lmodern}
\usepackage{amsmath,amssymb,amsfonts}
\usepackage{amsthm}
\usepackage{bm}
\usepackage{graphicx}
\usepackage{mathrsfs}
\usepackage{hyperref}
\usepackage{color}
\usepackage{slashed}
\usepackage{physics}
\usepackage{tensor}
\usepackage{booktabs}
\usepackage{multirow}

\usepackage{orcidlink}

\begin{document}

\title{On The Regularized McVittie Black Bounce and a Family of Traversable Wormholes}

\author{Soumya Chakrabarti \orcidlink{0000-0001-7255-9303} }
\email{soumya.chakrabarti@vit.ac.in}
\affiliation{
School of Advanced Sciences, Vellore Institute of Technology,
Tiruvalam Rd, Katpadi, Vellore, Tamil Nadu 632014
India
}
\author{Chiranjeeb Singha \orcidlink{0000-0003-0441-318X}}
\email{chiranjeeb.singha@iucaa.in}
\affiliation{Inter-University Centre for Astronomy and Astrophysics, Post Bag 4, Ganeshkhind, Pune - 411007, India}

\date{\today}

\begin{abstract}
The McVittie metric describes a self-gravitating compact object embedded in an expanding universe. It inherits two singularities: a curvature singularity at $r \to 0$ and a cosmological singularity at $a(t) \to 0$. Motivated by the construction of black-bounce geometries, we regularize both of these singularities and propose a regularized McVittie metric. The central singularity is regularized by replacing $r \to \sqrt{r^{2}+b^{2}}$, leading to a geometry that interpolates between a cosmological black hole, a black bounce, and a traversable wormhole. Similarly, the cosmological singularity is regularized by introducing a non-vanishing scale factor $a \rightarrow \sqrt{a^{2} + a_b^{2}}$. The resulting spacetime is supported by an effective imperfect fluid with finite anisotropic stress. We investigate the formation of the apparent horizon, analyze the circular geodesic structure, and argue that the geometry can also be interpreted as a conformally evolving Morris-Thorne wormhole embedded in a regular cosmological background.
\end{abstract}

\maketitle

\section{Introduction}
The gravitational interaction within our universe can be described by General Relativity (GR) with remarkable success. The cosmological description of the universe is one of the intriguing subtopics in GR which has received the most support from astrophysical observations. In this description, a cold dark matter governs the formation of large-scale structures, a candidate closely mimicking the cosmological constant drives a late-time acceleration, and their interplay determines the observed cosmic expansion \cite{planck1, planck2}. This description accounts for a wide range of observations, the cosmic microwave background, supernova distance measurements, and baryon acoustic oscillations \cite{obs1, obs2, obs3} to name a few. However, the description relies on fundamental ingredients whose physical origin remains elusive, such as the exact nature of dark matter, the identification of the mechanism responsible for primordial inflation, and the origin and equation of state (EOS) of dark energy \cite{tension1, tension2, tension3}. In addition, a notable tension remains among independent cosmological probes, particularly in their determination of the present-day Hubble expansion rate \cite{tension4, tension5, tension6}. Collectively, these issues point either to unresolved systematic effects or to new physics beyond the standard cosmological framework. In this context, non-singular cosmological scenarios deserve particular attention, where the initial big-bang singularity of the standard cosmological models is replaced by a regular bounce, connecting a preceding collapsing phase to the present expanding universe \cite{bounce1, bounce2, bounce3, bounce4, bounce5, bounce6, bounce7}.  \\

The idea of a non-singular `bouncing' cosmology motivates a number of questions concerning the fate of compact objects near the bounce. One can ask, for example, whether a black hole formed during the preceding collapsing phase can survive the bounce and persist into the subsequent epochs of expansion \cite{bhbounce1, bhbounce2}. There are examples of such primordial black holes which, apart from serving a novel mathematical curiosity, can also be used to derive implications for the resulting structure formation, dark matter distribution, and gravitational-wave related observations \cite{bhbounce3, bhbounce4, bhbounce5, bhbounce6}. Therefore, the study of black holes embedded in a cosmological background deserves particular attention. Among all such existing exact solutions describing a local inhomogeneity in an expanding universe, the McVittie metric holds a special place. Originally introduced by McVittie almost a century back \cite{mcvittie}, the metric describes a self-gravitating object embedded in a Friedmann-Lemaitre-Robertson-Walker (FLRW) universe. The properties and scope of this metric have been extensively studied in the literature, such as geometric and thermodynamic properties, observational signatures, and features of apparent horizons \cite{mcv1, mcv2, mcv3, mcv4}. \\

On the other hand, singularity-free compact geometries have received separate attention, particularly as an alternative to black holes and in the context of gravitational collapse \cite{nosing1, nosing2, nosing3, nosing4, nosing5, nosing6, nosing7, nosing8, nosing9}. Singularities are usually found near the core of a black hole \cite{penrose}, in GR as well as in its modifications \cite{fR}. They are interpreted as a general breakdown of classical theory rather than physical infinities. The possibility that this singular core can be replaced by a regular one has already inspired focus on exotic solutions such as wormholes, regular black holes, and black-bounces \cite{wh1, wh2, wh3, wh4, wh5, wh6, wh7, wh8, wh9}. A particularly elegant method of \textit{regularization} was recently introduced by Simpson and Visser \cite{simpsonvisser} by simply replacing the Schwarzschild radial coordinate $r \longrightarrow \sqrt{r^2+a^2}$, generating a one-parameter family of geometries that can interpolate between a Schwarzschild black hole, a regular black bounce, and a traversable wormhole. The resulting spacetime is regular for all coordinate ranges. Two questions can be asked in this context: through which process can such an astrophysically compact regular object form, and what should be the proper embedding of such an object? While the former question has received some attention, through models of gravitational collapse producing non-singular end-states \cite{scsk}, the latter remains unanswered. Astrophysical compact objects, although frequently modelled as residing in vacuum or radiating asymptotically flat geometries, are in fact expected to be immersed in an expanding cosmological background. This brings us to the question: can a wormhole/regular blackhole be embedded in a cosmological background, and if so, how does the presence of a regular core modify the horizon structure of such an object?   \\

The purpose of the present work is to address this question. We introduce a regularized McVittie metric by replacing the isotropic radial coordinate according to $r \longrightarrow s(r)\equiv\sqrt{r^{2}+b^{2}}$, where the parameter $b$ defines a regularization scale associated with the finite-area core. The resulting spacetime preserves the cosmological behaviour of the McVittie metric in an asymptotic limit, but replaces its central curvature singularity with a regular bounce surface. Depending on the parameter space, the geometry interpolates between a cosmological black hole, a black bounce and a traversable wormhole. We show that the corresponding matter source can be effectively interpreted as an anisotropic fluid violating the Null energy condition. We derive the conditions for the formation of the apparent horizon, study the circular geodesics governing the photon sphere, the Lyapunov exponent, and the innermost stable circular orbit. Finally, we extend the framework for regularization to the cosmological sector as well, by introducing a regularized scale factor. The completely regularized metric is free from a local as well as the background cosmological singularity. \\

The paper is organized as follows. In Section \ref{geometry}, we describe the geometry of the regularized McVittie spacetime. Section \ref{fieldequation} presents the Einstein field equations and the components of the effective imperfect fluid supporting the spacetime. In Section \ref{apparenthorizon}, we determine the location of the apparent horizon. Section \ref{geodesic} is devoted to the analysis of circular geodesics, the photon sphere, the Lyapunov exponent, and the innermost stable circular orbit (ISCO). In Section \ref{scalefactor}, we regularize the cosmological scale factor. Finally, in Section \ref{Conclusion}, we summarize our main findings and discuss their possible implications. \\

\textit{Notations and Conventions:} Throughout this work, we use the $\mathrm{diag}(-1,+1,+1,+1)$ convention. Unless otherwise specified, all calculations are performed in natural units where the gravitational constant (G) and the speed of light (C) are set to unity, i.e., $G=C=1$.

\section{Geometry of the Regularized McVittie Spacetime}\label{geometry}
The McVittie solution describes a black hole embedded in an expanding FLRW universe. Unlike the standard, asymptotically flat black-hole metrics, it incorporates the effects of cosmological expansion while preserving spherical symmetry. The original solution was derived under the assumption that the central object does not accrete the cosmological background, which effectively leads to a constant McVittie mass parameter \cite{mcvittie, mcv1}. Relaxing the no-accretion condition leads to different forms of a generalized McVittie, and such modifications have since been proposed (see for instance \cite{faraonimcv}). However, the original solution remains a canonical starting point if one wants to construct a compact object embedded in a cosmological background. In isotropic coordinates, the McVittie metric is written as

\begin{equation}
d\bar{s}^2 = -\left(\frac{1-\mu}{1+\mu}\right)^2dt^2 + a^2(t)(1+\mu)^4 \left(dr^2 + r^2d\Omega^2\right),
\label{McVittieMetric}
\end{equation}

where

\begin{equation}
\mu(t,r)=\frac{M}{2a(t)r}.
\label{muMcVittie}
\end{equation}

$M$ is the McVittie mass parameter and $a(t)$ denotes the cosmological scale factor. In the $r \rightarrow \infty$ limit, $\mu$ vanishes, and the metric reduces to a spatially flat FLRW geometry. On the other hand, for $a(t) = 1$, the metric reduces to a Schwarzschild geometry. One of the remarkable features of the McVittie metric is the nontrivial structure of its effective matter source: the energy density is spatially homogeneous (coincides with the background FLRW density) while the pressure acquires inhomogeneity. This distinct feature is in contrast with conventional compact objects and has motivated investigations regarding the corresponding equation of state \cite{harada}. It has also been proven that the original McVittie metric cannot, in general, be supported by a single minimally coupled canonical scalar field \cite{abdalla, afshordi, nolan}. There are also recent reports on the violation of some of the energy conditions sufficiently close to the singular core of McVittie \cite{necmcvittie}. \\

While we acknowledge the existing features of McVittie, we remain particularly focused on the central curvature singularity associated with the Schwarzschild-like core. Motivated by the need for cosmologically embedded regular compact objects, we attempt a regularization analogous to the Simpson-Visser construction \cite{simpsonvisser} within the McVittie framework. We introduce a regularized radial coordinate as

\begin{equation}
r \longrightarrow s(r) \equiv \sqrt{r^2+b^2},
\label{Regularization}
\end{equation}

where $b > 0$ is a constant length scale. The resulting spacetime is then written as

\begin{equation}
d\bar{s}^2 = -\left(\frac{1-\mu}{1+\mu}\right)^2dt^2 + a^2(t)(1+\mu)^4 \left[dr^2 + s(r)^2 d\Omega^2 \right],
\label{RegMcVittie}
\end{equation}

with

\begin{equation}
\mu(t,r) = \frac{M}{2a(t)\sqrt{r^2+b^2}} = \frac{M}{2a(t)s}.
\label{muRegularized}
\end{equation}

The radial coordinate $r$ falls within the range $-\infty < r < +\infty$. Unlike the ordinary McVittie geometry, no curvature singularity occurs at $r = 0$. We rewrite the regularized metric using the modified radial coordinate $s(r) = \sqrt{r^2+b^2}$ as
\begin{equation}
d\bar{s}^{2} = -\left(\frac{1-\mu}{1+\mu}\right)^{2}dt^{2} + a^{2}(t)(1+\mu)^{4}\left[\frac{ds^{2}}{1 - \frac{b^{2}}{s^2}} + s^{2}d\Omega^{2}
\right],
\label{Metriclcoordinate}
\end{equation}

where
\begin{equation}
r = \pm\sqrt{s^{2}-b^{2}} ~,~ s \ge b ~,~ \mu(t,s)=\frac{M}{2a(t)s}.
\end{equation}

Unlike the isotropic coordinate $r$, the new radial coordinate is restricted by $s \ge b$. As a result, the two FLRW regions located at $r \rightarrow \pm\infty$ are now connected through a two-sphere of finite area $A_{\rm min}=4\pi b^{2}$. We interpret this as a geometric throat of the spacetime, since Eq. (\ref{Metriclcoordinate}) bears a striking resemblance to the canonical Morris-Thorne wormhole metric \cite{wh1, wh2}, with the factor $(1 - \frac{b^{2}}{s^2})^{-1}$ playing the role of a shape function. For a spherically symmetric wormhole, a throat is identified as a closed two-surface of minimum area, which is equivalent to the areal radius reaching a local minima. We study the extrema of the areal radius, defined as
\begin{equation}
Y(t,r) = a(t)(1+\mu)^2 \sqrt{r^2+b^2} = a(t)(1+\mu)^2 s.
\label{ArealRadius}
\end{equation}

Differentiating with respect to $r$, we find that 
\begin{equation}
\left. \frac{\partial Y}{\partial r} \right|_{r=0} = \left. \frac{a r}{s} (1+\mu)(1-\mu)\right|_{r=0} = 0.
\label{drdR}
\end{equation}

Therefore $r = 0$ corresponds to an extremal surface. The areal radius of the circle at this surface is

\begin{equation}
Y_{\rm th}(t) = ab \left(1+\frac{M}{2ab}\right)^2,
\label{ThroatRadius}
\end{equation}

which is strictly positive, independent of the choice of parameters. Thus the central singularity of the standard McVittie metric is replaced by a finite-area surface that acts as a throat joining two asymptotic cosmological regions. The second derivative of the areal radius is

\begin{eqnarray}\label{secondthroat}
&& \left. \frac{\partial^2 Y}{\partial r^2} \right|_{r=0} = \frac{a}{b}\left(1 + \frac{M}{2ab}\right)\left(1 - \frac{M}{2ab}\right), \\&&\label{secondthroat1}
\frac{M}{2ab} < 1 \qquad \Longrightarrow \qquad \frac{\partial^2 Y}{\partial r^2} > 0,\\&&\label{secondthroat2}
\frac{M}{2ab} > 1 \qquad \Longrightarrow \qquad \frac{\partial^2 Y}{\partial r^2} < 0.
\end{eqnarray}
While the condition in Eq. (\ref{secondthroat1}) corresponds to a traversable wormhole throat, Eq. (\ref{secondthroat2}) corresponds to a black-bounce, analogous to the Simpson-Visser construction. The condition $\partial^2 Y/\partial r^2 > 0$ can be seen as the dynamical analogue of the familiar flare-out condition for Morris-Thorne wormholes.  \\  

We can also prove that the areal radius of the extremal surface as in Eq. (\ref{ThroatRadius}) has a unique global minimum with respect to the dimensionless parameter $ab$. This minimum occurs at the critical value $2ab = M$, which coincides with the transition separating the traversable wormhole and black-bounce branches of the geometry. Expanding the throat radius gives
\begin{equation}
Y_{\rm th} = ab+M+\frac{M^2}{4ab}.
\end{equation}
Differentiating with respect to $ab$ gives
\begin{equation}
\frac{d Y_{\rm th}}{d(ab)} = 1 - \frac{M^2}{4(ab)^2}.
\end{equation}
The stationary point therefore satisfies
\begin{equation}
\frac{d Y_{\rm th}}{d(ab)} = 0 \qquad \Longrightarrow \qquad 2ab = M.
\end{equation}
Furthermore,
\begin{equation}
\frac{d^2 Y_{\rm th}}{d(ab)^2} = \frac{M^2}{2(ab)^3}>0,
\end{equation}
showing that the stationary point is a global minima.  \\

The lapse function and the behaviour of radial null geodesics provide an independent characterization of the transition between a wormhole and a black-bounce. We note that the lapse function $g_{tt} = - \left(\frac{1-\mu}{1+\mu}\right)^2$ can only vanish when $1-\mu = 0$, or equivalently, at $s = \frac{M}{2a}$. Because the regularized radial coordinate satisfies $s \ge b$, this surface can exist for $2ab \le M$ alone. This is precisely the critical condition that separates a traversable wormhole from a black-bounce branch.  \\

To explore the radial null curves, we set $d\Omega = 0$, $ds^2 = 0$ and write
\begin{equation}
\frac{dr}{dt} = \pm \frac{1-\mu}{a(1+\mu)^3},
\end{equation}
where the positive and negative signs correspond to outgoing and ingoing null rays, respectively. The coordinate velocity depends on $\mu = M/(2as)$. For $2ab > M$, one has $\mu < 1$ everywhere, since $s \ge b$. Therefore, $\frac{dr}{dt}$ is never zero, i.e., radial null rays can propagate across the bounce surface and connect the two asymptotically cosmological regions. At the critical value $2ab = M$, the minimum admissible radius satisfies $s = b = \frac{M}{2a}$, so that $g_{tt} = 0$. Finally, when $2ab<M$, the equation $\mu=1$ admits the two real solutions
\begin{equation}
r_{\rm L} = \pm \sqrt{\frac{M^2}{4a^2}-b^2},
\end{equation}
which correspond to the hypersurfaces where the lapse function vanishes. Across these hypersurfaces $dr/dt\rightarrow0$, signalling the formation of horizon-like surfaces that conceal the regular bounce from distant observers.    \\

It is important to emphasize that the surfaces defined by $g_{tt} = 0$ are not, in general, identical to the apparent horizons. The former characterize the behaviour of the causal nature of the coordinate system, whereas the apparent horizons are determined by the vanishing of the expansion of outgoing null congruences. The two notions coincide only in a static spacetime, which a McVittie geometry is not! 

\subsection{Regularity of Curvature Scalars}
A necessary follow-up is to confirm the absence of curvature singularities. In order to do that we first note that $s(r)=\sqrt{r^2+b^2} \ge b > 0$. Therefore, $s(r)$, $\frac{ds}{dr}$, $\mu$, $\frac{\partial \mu}{\partial r}$ and $\frac{\partial^{2} \mu}{\partial r^{2}}$ remain bounded everywhere. Therefore, the metric coefficients, their derivatives, curvature tensor components, and the scalar invariants remain continuous and bounded for every finite value of $r$. The Ricci scalar can be derived as

\begin{figure}[h]
	\begin{center}
		\includegraphics[angle=0, width=0.35\textwidth]{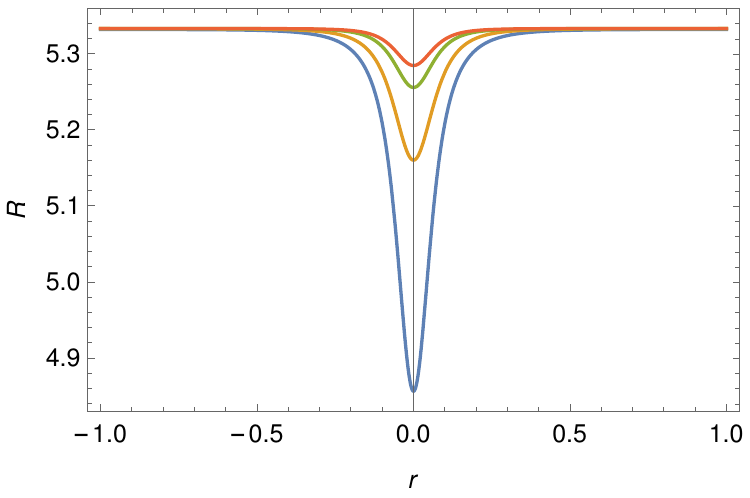}
		\includegraphics[angle=0, width=0.35\textwidth]{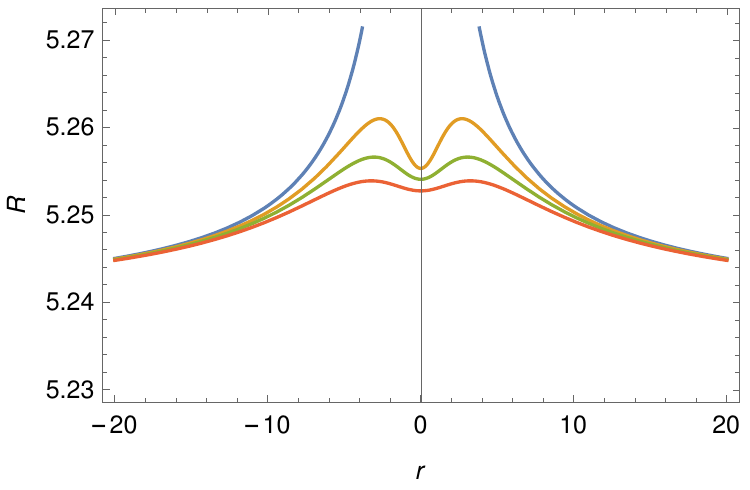}
	\end{center}
	\caption{{\small Ricci scalar as a function of $r$. The top graph shows the profile with different values of time $t$ (considered as a parameter) and graph below shows the profile with different values of the regularization parameter $b$.}}
	\label{Ricci_1}
\end{figure}

\begin{widetext}
\footnotesize
\begin{eqnarray}\nonumber
&&\mathcal{R} = \Bigg[2 \Big(M^2-4 a^2 \Big(b^2+r^2\Big)\Big) \Big(192 a^7 \Big(b^2+r^2\Big)^{7/2} \ddot{a}-9 M^6 \sqrt{b^2+r^2} \dot{a}^2+3 M^5 \Big(M \sqrt{b^2+r^2} \ddot{a} -28 \Big(b^2+r^2\Big) \dot{a}^2\Big)\\&&\nonumber
+12 M^4 a^2 \Big(b^2+r^2\Big) \Big(3 M \ddot{a} - 25 \sqrt{b^2+r^2} \dot{a}^2 \Big)-60 M^3 a^3 \Big(b^2+r^2\Big) \Big(8 \Big(b^2+r^2\Big) \dot{a}^2-3 M \sqrt{b^2+r^2} \ddot{a}\Big)\\&&\nonumber
-48 M^2 a^4 \Big(-10 M \Big(b^2+r^2\Big)^2 \ddot{a} + 5 \Big(b^2+r^2\Big)^{5/2} \dot{a}^2+b^2 \sqrt{b^2+r^2}\Big)+64 a^6 \Big(b^2+r^2\Big) \Big(9 M \Big(b^2+r^2\Big)^2 \ddot{a} \\&&\nonumber
+3 \Big(b^2+r^2\Big)^{5/2} \dot{a}^2-b^2 \sqrt{b^2+r^2}\Big)+16 M a^5 \Big(b^2+r^2\Big) \Big(45 M \Big(b^2+r^2\Big)^{3/2} \ddot{a} + 12 \Big(b^2+r^2\Big)^2 \dot{a}^2+4 b^2\Big)\Big)\Big]\\&&\label{ricciiiii}
\Big[a^2 \sqrt{b^2+r^2} \Big(M-2 a \sqrt{b^2+r^2}\Big)^2 \Big(2 a \sqrt{b^2+r^2}+M\Big)^6\Big]^{-1}.
\end{eqnarray}
\end{widetext}

From the above equation, we show that the Ricci scalar is inversely proportional to the factor $a^2 \sqrt{b^2+r^2} \Big(M-2 a \sqrt{b^2+r^2}\Big)^2 \Big(2 a \sqrt{b^2+r^2}+M\Big)^6$, which never goes to zero at any real value of $r$. Thus, the curvature singularity at $r \rightarrow 0$ is removed due to the regularization. We confirm this by plotting Eq. (\ref{ricciiiii}) for a scale factor chosen at the outset ($a(t) \sim a_0 e^{H_0 t}$). We plot as a function of $r$, allowing variation of two different parameters, namely, the time $t$ (top graph) and the regularization parameter $b$ (bottom graph) in FIG. \ref{Ricci_1}. We also emphasize that the spacetime continues to inherit any cosmological singularities that might be associated with the background FLRW scale factor $a(t)$. In particular, if $a(t) \rightarrow 0$, at any finite cosmic time, the spacetime will continue to develop a global cosmological singularity. This singularity originates in the background cosmology rather than in the local compact-object geometry and we defer its discussion and subsequent resolution towards the latter part of the article.  

\section{Einstein Field Equations and an Effective Imperfect Fluid}\label{fieldequation}
We investigate whether the matter content supporting the regularized McVittie geometry can be described by an effective imperfect fluid. First, we rewrite the metric in a generic form

\begin{eqnarray}\label{SphMetric}
&&d\bar{s}^2 = -A^2(t,r) dt^2 + B^2(t,r) dr^2 + Y^2(t,r)d\Omega^2 ,\nonumber\\\\&&\label{Adef}
A(t,r) = \frac{1-\mu}{1+\mu} ~~,~~ B(t,r) = a(t)(1+\mu)^2, \\&&\label{rdef}
Y(t,r) = a(t)(1+\mu)^2 s(r), s(r)=\sqrt{r^2+b^2}, \\&&\label{mudef}
\mu(t,r)=\frac{M(t)}{2a(t)s}.
\end{eqnarray}

For the time being, we have kept $M$ as a function of time. The nonvanishing mixed components of the Einstein tensor can be derived as

\begin{eqnarray}\nonumber\label{Gtt}
&& G^{t}{}_{t} = -\frac{2}{Y B^2} \left(Y'' -\frac{B'}{B}Y' \right) - \frac{1}{Y^2} \left[ 1-\frac{Y'^2}{B^2} + \frac{\dot Y^{\,2}}{A^2} \right]
\\&&
-\frac{2}{A^2}\frac{\ddot Y}{Y} + \frac{2}{A^2}\frac{\dot A}{A}\frac{\dot Y}{Y}, \\&&
G^{t}{}_{r} = \frac{2}{Y}\left(\frac{\dot Y'}{A^2} - \frac{A' \dot Y}{A^3} - \frac{r'\dot{B}}{A^2 B} \right), \\&&\nonumber\label{GtR}
G^{r}{}_{r} = -\frac{2}{YA^2} \left(\ddot Y - \frac{\dot A}{A}\dot Y \right) - \frac{1}{Y^2} \left[ 1-\frac{Y'^2}{B^2} + \frac{\dot Y^{\,2}}{A^2} \right] \\&&
+ \frac{2}{B^2} \frac{A'}{A} \frac{Y'}{Y},\\&&\nonumber\label{GRR}
G^{\theta}{}_{\theta} = G^{\phi}{}_{\phi} = -\frac{1}{YA^2}\left( \ddot Y - \frac{\dot A}{A}\dot Y \right) \\&&+ \frac{1}{YB^2} \left( Y'' - \frac{B'}{B}Y' \right)
- \frac{1}{2Y^2} \left[1-\frac{Y'^2}{B^2}+\frac{\dot Y^{\,2}}{A^2} \right].\label{Gthth}
\end{eqnarray}

The $G^{t}{}_{R}$ component is derived as
\begin{equation}
G^{t}{}_{r} = \frac{2r\,\dot M} {a^{2}s^{3}(1-\mu)^{2}(1+\mu)^{5}},
\label{GtR_simplified}
\end{equation}
where $\dot M \equiv dM/dt$. The generic energy-momentum tensor for an imperfect fluid can be written as

\begin{equation}
T_{\mu\nu} = (\rho+p_t)u_\mu u_\nu + p_t g_{\mu\nu} + (p_r-p_t)\chi_\mu\chi_\nu + q_{(\mu}u_{\nu)},
\label{EMtensor}
\end{equation}

where

\begin{eqnarray}
&& u^\mu = \frac{1+\mu}{1-\mu} \delta^\mu_t ~~,~~ \chi^\mu = \frac{1}{a(1+\mu)^2} \delta^\mu_r ,\\&&
u^\mu u_\mu = -1 ~~,~~ \chi^\mu\chi_\mu = 1 ~~,~~ u^\mu\chi_\mu = 0.
\end{eqnarray}

The mixed component of the Einstein tensor becomes zero when the mass parameter is treated as a constant, i.e., for the McVittie no-accretion condition. One has two options regarding this: (i) to keep the dynamical mass parameter and include a non-zero radial heat flux, $T^{t}{}_{R} \neq 0$ to support it, or (ii) treat the mass parameter as a constant and describe the interior as a non-radiating anisotropic fluid. We explore the latter, for which the heat flux disappears and the effective fluid components can be related to the geometry as

\begin{equation}\label{EMTdescr}
\rho = -\frac{1}{8\pi}G^{t}{}_{t},~ q^r = \frac{1}{8\pi A} G^{t}{}_{r},~ p_r = \frac{1}{8\pi}G^{r}{}_{r},~ p_t = \frac{1}{8\pi} G^{\theta}{}_{\theta}.
\end{equation}

There is an important distinction in the regularized McVittie metric. In the original McVittie geometry, the no-accretion condition leads to an energy density that depends only on time, $\rho = \rho(t)$, whereas the pressure is spatially inhomogeneous, $p = p(t,r)$. This is a rather restrictive feature and makes the physical interpretation of the matter source non-trivial. The regularization $r\rightarrow s(r)=\sqrt{r^2+b^2}$ resolves the issue. For $b\neq0$, the effective energy density becomes spatially inhomogeneous, and the radial and tangential pressures remain inhomogeneous and, in general, distinct from one another. 

\section{Expansion of the Radial Null Congruences and Energy Conditions}\label{energycondition}
The causal structure of a spherically symmetric spacetime can be characterized by the expansions of outgoing and ingoing radial null
congruences. We take a radial null vector in the form
\begin{equation}
k^\mu=(k^t,k^r,0,0) ~~,~~ g_{\mu\nu}k^\mu k^\nu=0.
\end{equation}
For the regularized McVittie metric as in Eq. (\ref{SphMetric}) this immediately gives $k^r=\pm\frac{A}{B}k^t$. We choose the normalization of the null vector as $k^t=\frac{1}{A}$ and write the future-directed outgoing and ingoing radial null vectors as
\begin{equation}
\ell^\mu = \left(\frac1A,\frac1B,0,0 \right) ~,~ n^\mu = \left(\frac{1}{A}, -\frac{1}{B}, 0, 0 \right).
\end{equation}
The expansion of a null congruence generated by a vector field $k^\mu$ is derived from
\begin{equation}
\theta = \frac{2}{Y} k^\mu\nabla_\mu Y.
\end{equation}

For the outgoing congruence,
\begin{equation}
\ell^\mu\nabla_\mu Y = \frac{\dot Y}{A} + \frac{Y'}B \Rightarrow \theta_+ = \frac{2}{Y}\left(\frac{\dot Y}{A} + \frac{Y'}B\right).
\end{equation}

Similarly, for the ingoing congruence,
\begin{equation}
n^\mu\nabla_\mu Y = \frac{\dot Y}{A} - \frac{Y'}B \Rightarrow \theta_- = \frac{2}{Y}\left(\frac{\dot Y}{A} - \frac{Y'}B \right).
\end{equation}

Using the metric coefficients as in Eq. (\ref{SphMetric}), and the condition $\dot{M} = 0$, the null expansions can be simplified into
\begin{equation}\label{rcrc}
\theta_\pm = 2H \pm \frac{2r(1-\mu)}{a(1+\mu)^3s^2}.
\end{equation}

The product of the two expansions is written as
\begin{equation}\label{thetatheta}
\theta_+\theta_- = -\frac4{Y^2}g^{ab}\nabla_a Y\nabla_b Y = \frac{4}{Y^2} \left(\frac{\dot Y^2}{A^2} - \frac{Y'^2}{B^2}\right).
\end{equation}

At the bounce surface $r = 0$, we find that $\theta_{+}(0)=\theta_{-}(0)=2H$, which suggests that at this point the local expansion of null congruences is governed purely by the background cosmological expansion. If $H > 0$, both congruences initially move away from one another around the regularized throat, whereas for $H < 0$, both seem to converge. This behaviour is in contrast with that of static wormhole geometries, where the null expansions vanish identically at the throat.  \\

The evolution of the null expansions is governed by the Raychaudhuri equation \cite{rc}. For a congruence of future-directed null geodesics with tangent vector $k^\mu$, the Raychaudhuri equation reads
\begin{equation}
\frac{d\theta}{d\lambda} = -\frac{1}{2}\theta^2 - \sigma_{\mu\nu}\sigma^{\mu\nu} + \omega_{\mu\nu}\omega^{\mu\nu} - R_{\mu\nu}k^\mu k^\nu,
\end{equation}
where $\lambda$ is an affine parameter, $\sigma_{\mu\nu}$ is the shear tensor and $\omega_{\mu\nu}$ is the rotation tensor. For the radial null congruences considered here, spherical symmetry implies that the congruence is hypersurface orthogonal and therefore $\omega_{\mu\nu} = 0$. Furthermore, the absence of any preferred angular direction forces the shear to vanish, i.e., $\sigma_{\mu\nu} = 0$. The Raychaudhuri equation, therefore, reduces to
\begin{equation}
\frac{d\theta_\pm}{d\lambda} = -\frac{1}{2}\theta_\pm^2 - R_{\mu\nu}k^\mu_\pm k^\nu_\pm .
\end{equation}

Using the Einstein field equations and the null condition $g_{\mu\nu}k^\mu k^\nu = 0$, one can write the Ricci term as $R_{\mu\nu}k^\mu k^\nu = 8\pi T_{\mu\nu}k^\mu k^\nu$. For an effective imperfect fluid with heat flux $q$,
\begin{equation}
T_{\mu\nu} = (\rho+p_t)u_\mu u_\nu + p_tg_{\mu\nu} + (p_r-p_t)\chi_\mu\chi_\nu + q_{(\mu}u_{\nu)},
\end{equation}
the outgoing and ingoing null vectors satisfy
\begin{equation}
u_\mu k^\mu_\pm=-1 ~~,~~ \chi_\mu k^\mu_\pm= \pm 1.
\end{equation}
It follows immediately that
\begin{equation}
T_{\mu\nu}k^\mu_\pm k^\nu_\pm = \rho + p_r \pm 2q.
\end{equation}
The Raychaudhuri equation, therefore, assumes the form
\begin{equation}
\frac{d\theta_\pm}{d\lambda} = -\frac{1}{2}\theta_\pm^2 - 8\pi (\rho+p_r\pm2q).
\end{equation}

This equation provides the geometric interpretation of the Null Convergence Condition \cite{ncc} or the Null Energy Condition (NEC) \cite{nec1, nec2} in GR. For all $\rho + p_r \pm 2q \ge 0$, the second term on the right-hand side is negative, which causes adjacent null geodesics to focus. On the other hand, violation of this condition leads to the defocusing of radial null congruences. The latter of these conditions constitutes the geometric origin of the flare-out condition around wormhole throats. If the McVittie mass parameter is independent of time, i.e., the radial heat flux is zero, the Raychaudhuri equation is simplified into
\begin{equation}
\frac{d\theta_\pm}{d\lambda} = -\frac{1}{2}\theta_\pm^2 - 8\pi(\rho+p_r).
\end{equation}

\begin{figure}[h]
	\begin{center}
		\includegraphics[angle=0, width=0.35\textwidth]{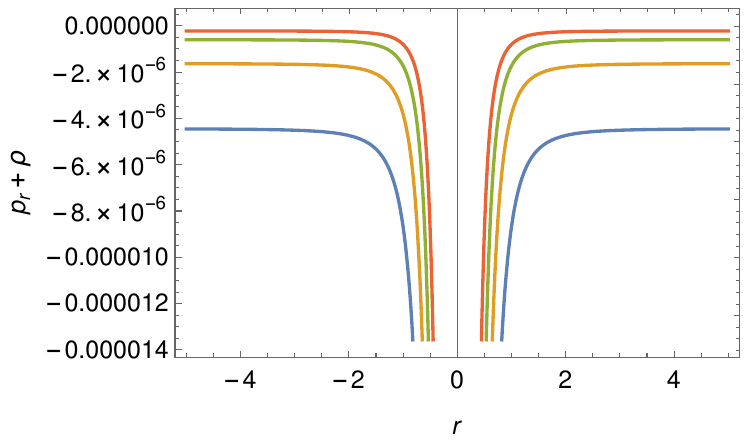}
		\includegraphics[angle=0, width=0.35\textwidth]{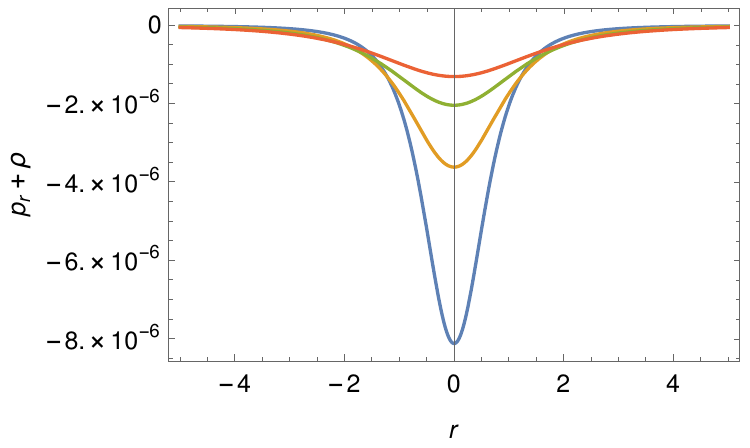}
		\includegraphics[angle=0, width=0.35\textwidth]{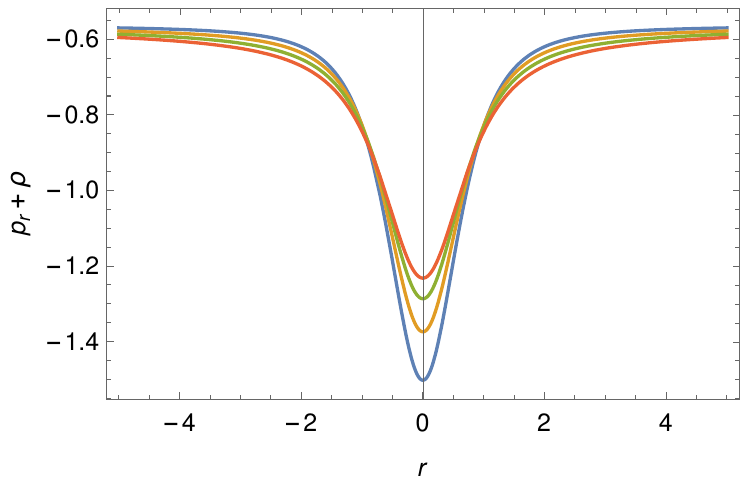}
		\includegraphics[angle=0, width=0.35\textwidth]{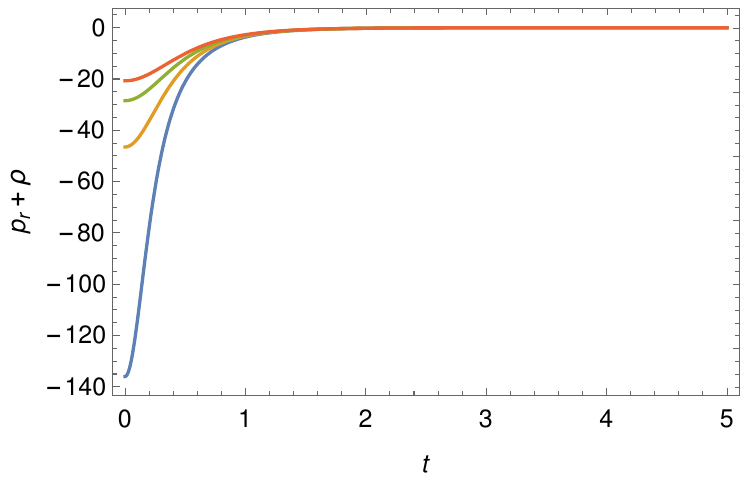}
	\end{center}
	\caption{{\small The profile of left-hand side of NEC, $\rho + p_r$, for four different setups. Graph on top : $\rho + p_r$ as a function of $r$, for different values of coordinate time $t$. Graph second from top : $\rho + p_r$ as a function of $r$, for different values of the regularization parameter $b$. Graph third from top : $\rho + p_r$ as a function of $r$ for different values of the McVittie mass $M$. Graph below : $\rho + p_r$ as a function of time, for different shells, i.e., different values of $r$.}}
	\label{NEC_1}
\end{figure}

\begin{figure}[h]
	\begin{center}
		\includegraphics[angle=0, width=0.35\textwidth]{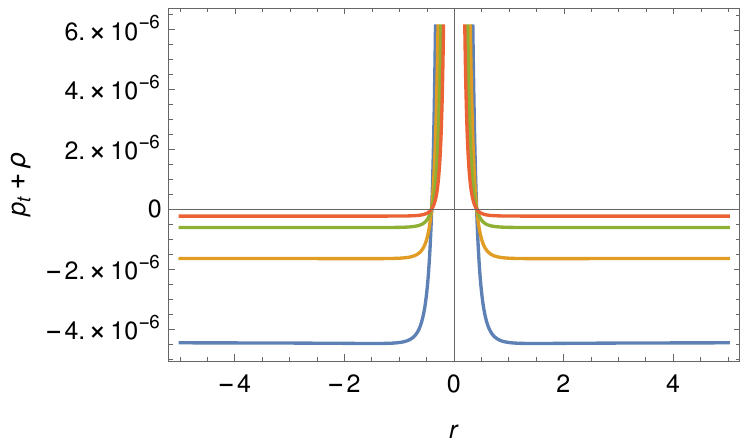}
		\includegraphics[angle=0, width=0.35\textwidth]{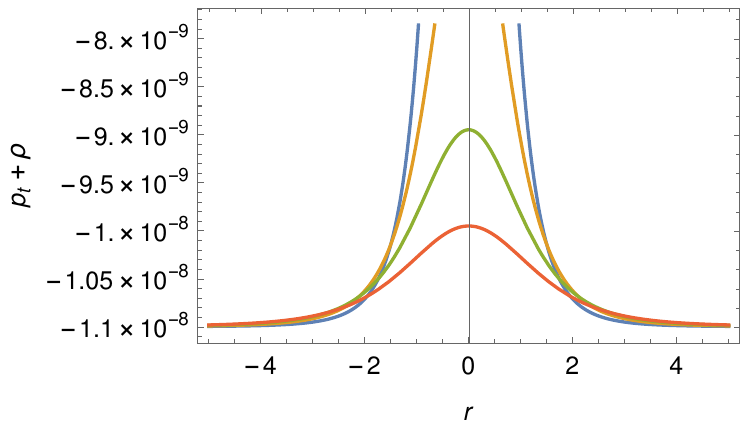}
		\includegraphics[angle=0, width=0.35\textwidth]{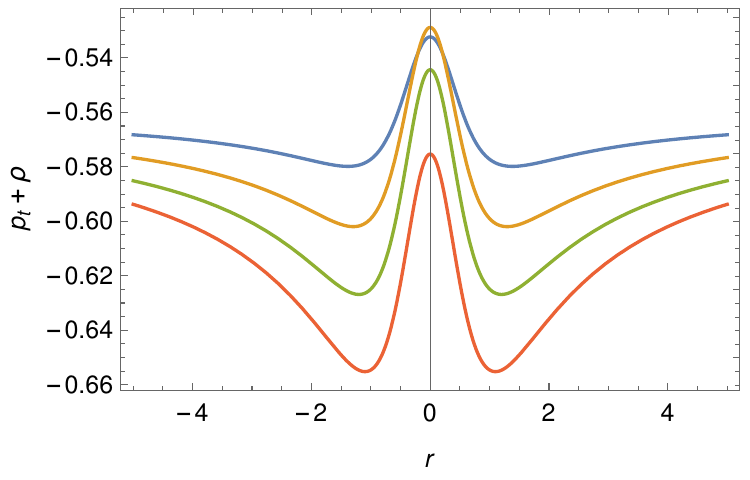}
		\includegraphics[angle=0, width=0.35\textwidth]{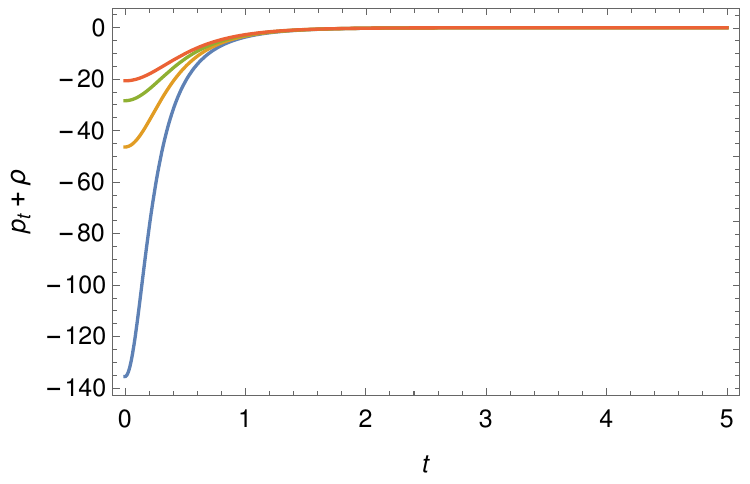}
	\end{center}
	\caption{{\small The profile of left-hand side of WEC, $\rho + p_t$, for four different setups. Graph on top : $\rho + p_t$ as a function of $r$, for different values of coordinate time $t$. Graph second from top : $\rho + p_t$ as a function of $r$, for different values of the regularization parameter $b$. Graph third from top : $\rho + p_t$ as a function of $r$ for different values of the McVittie mass $M$. Graph below : $\rho + p_t$ as a function of time, for different shells, i.e., different values of $r$.}}
	\label{WEC_1}
\end{figure}

Apart from NEC, in general, there are other identities treated as energy conditions. These are derived from an eigenvalue problem of the generic energy-momentum tensor \cite{nec2}, and they ensure a strict positivity of the energy density. For an arbitrary stress-energy tensor $T_{\mu\nu}$, apart from NEC the three other energy conditions are defined as follows (i) Weak Energy Condition (WEC) which requires that for every timelike vector $u^\mu$, $T_{\mu\nu}u^\mu u^\nu \ge 0$ ; (ii) Dominant Energy Condition (DEC), which requires, in addition to WEC that for every future-directed timelike vector $u^\mu$, $T^{\mu}{}_{\nu}u^\nu$ is non-spacelike and (iii) Strong Energy Condition (SEC) which requires that for every timelike vector $u^\mu$, $\left(T_{\mu\nu}-\frac{1}{2} T g_{\mu\nu} \right) u^\mu u^\nu \ge0$. For an anisotropic fluid as written in Eq. (\ref{EMtensor}), these energy conditions simply become

\begin{eqnarray}\label{NEC}
&& NEC \Rightarrow \rho+p_r \ge 0~~,~~\rho+p_t \ge 0,\\&&\label{WEC}
WEC \Rightarrow \rho \ge 0 ~~,~~ \rho+p_r \ge 0 ~~,~~ \rho+p_t \ge 0,\\&&\label{DEC}
DEC \Rightarrow \rho \ge |p_r|~~,~~ \rho \ge |p_t|,\\&&\label{SEC}
SEC \Rightarrow \rho + p_r + 2p_t \ge 0.
\end{eqnarray}

Using the field equations (Eqs. (\ref{Gtt})-(\ref{Gthth})), it is straightforward to derive the full expressions of these energy conditions.
For a toy scale factor chosen at the outset, $a(t) \sim a_0 e^{H_0 t}$, we plot them. In FIG. \ref{NEC_1}, we plot the left-hand side of NEC, i.e., $\rho + p_r$ for four different setups. First, we plot it as a function of $r$, for different values of coordinate time $t$ (graph on top). We also plot the NEC for different values of the regularization parameter $b$ (graph second from the top) and the McVittie mass $M$ (graph third from the top). Finally, we plot the NEC as a function of time, for different shells, i.e., different values of $r$ (graph below). For all the cases considered, it can be seen that there is a general violation of the Null Energy Condition.  \\

\begin{figure}[h]
	\begin{center}
		\includegraphics[angle=0, width=0.35\textwidth]{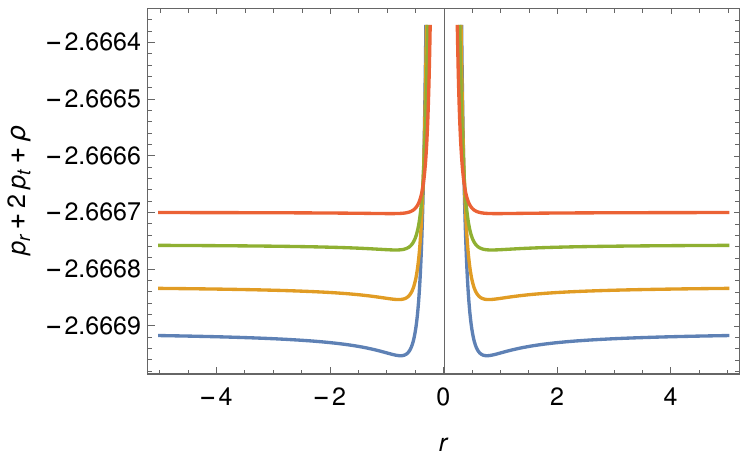}
		\includegraphics[angle=0, width=0.35\textwidth]{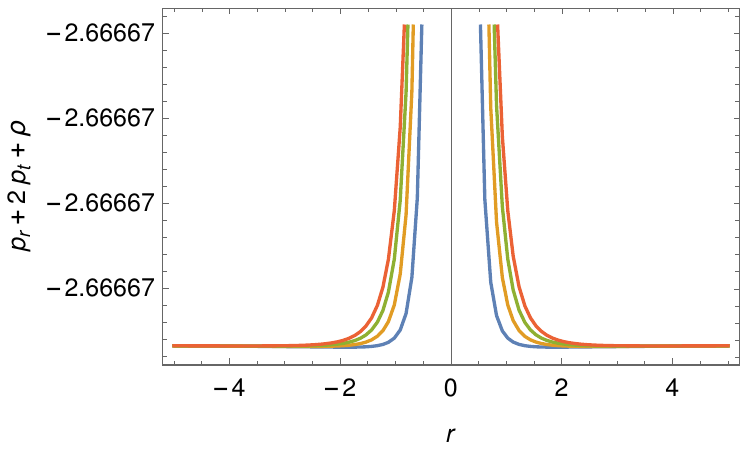}
		\includegraphics[angle=0, width=0.35\textwidth]{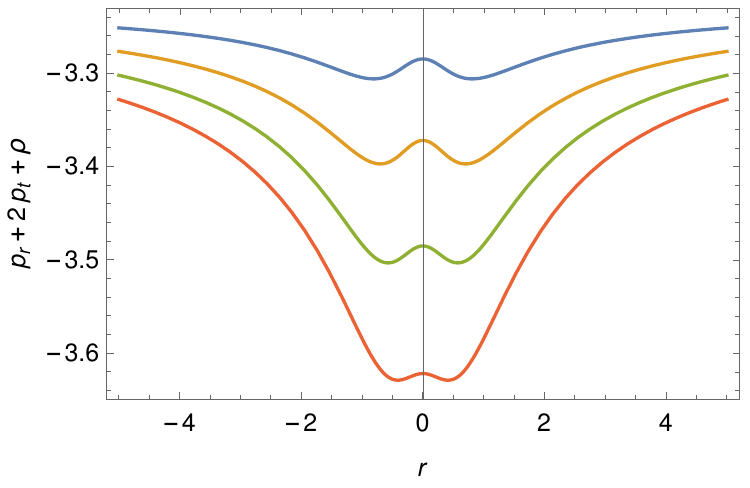}
		\includegraphics[angle=0, width=0.35\textwidth]{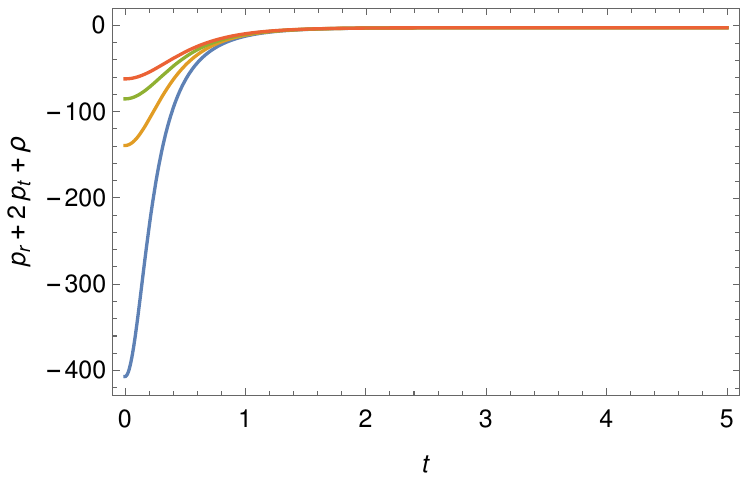}
	\end{center}
	\caption{{\small The profile of left-hand side of SEC, $\rho + p_r + 2p_t$, for four different setups. Graph on top : $\rho + p_r + 2p_t$ as a function of $r$, for different values of coordinate time $t$. Graph second from top : $\rho + p_r + 2p_t$ as a function of $r$, for different values of the regularization parameter $b$. Graph third from top : $\rho + p_r + 2p_t$ as a function of $r$ for different values of the McVittie mass $M$. Graph below : $\rho + p_r + 2p_t$ as a function of time, for different shells, i.e., different values of $r$.}}
	\label{SEC_1}
\end{figure}

We find that the violation of energy conditions is a generic pattern because of the regularization at the core. The WEC, as well as the SEC, shows a similar signature of this violation, as can be seen from FIG. \ref{WEC_1} and FIG. \ref{SEC_1}.

\section{Location of Apparent Horizon(s)}\label{apparenthorizon}
In dynamical spacetimes, an apparent horizon is characterized by the vanishing of one of the null expansions. This is characterized by Eq. (\ref{thetatheta}),
\begin{equation}\label{AHcondition}
\theta_+\theta_- = 0 \qquad\Longleftrightarrow\qquad g^{ab}\nabla_a Y \nabla_b Y = 0.
\end{equation}
The areal radius $Y(t,r)$ is given as
\begin{equation}
Y(t,r) = a(t)(1+\mu)^2 s \qquad s(r) = \sqrt{r^2+b^2}.
\label{ArealRadiusAgain}
\end{equation}

Using the metric coefficients, we simplify Eq.~(\ref{AHcondition}) into
\begin{equation}
\frac{r_{ap}^2}{r_{ap}^2+b^2} = \frac{(1+\mu)^6}{(1-\mu)^4} \left[ H a (1-\mu) s_{ap} + \dot M \right]^2 .
\label{GeneralHorizonEq}
\end{equation}

Eq. (\ref{GeneralHorizonEq}) determines the coordinate locations of the apparent horizon(s) of the regularized McVittie spacetime. $r_{ap}$ and $s_{ap}$ are associated with the modified radial coordinate associated with the metric. Using $\dot M = 0$ we simplify the equation into
\begin{equation}
\frac{r_{ap}^2}{r_{ap}^2+b^2} = 1-\frac{b^2}{s_{ap}^2} = H^2 Y_{ap}^2 \frac{(1+\mu)^2}{(1-\mu)^2}.
\label{HorizonEq1}
\end{equation}

We note that the areal radius $Y = a\left(1+\frac{M}{2as}\right)^2 s$ can be converted into a quadratic as 
\begin{equation}\label{quadratic}
as^2+(M-Y)s+\frac{M^2}{4a}=0,
\end{equation}
solving which we get
\begin{equation}
s = \frac{Y - M + \sqrt{Y(Y-2M)}}{2a}.
\label{sintermsofr}
\end{equation}
Using Eq. (\ref{sintermsofr}), the parameter $\mu=\frac{M}{2as}$ can be expressed in terms of the areal radius as
\begin{equation}\label{LapseRatio}
\mu = \frac{M}{Y-M+\sqrt{Y(Y-2M)}}.
\end{equation}

Substituting Eqs. (\ref{sintermsofr}) and (\ref{LapseRatio}) into Eq.~(\ref{HorizonEq1}) we write the equation governing the location of apparent horizon, in terms of the physical areal radius as
\begin{eqnarray}\nonumber
&& 1 - H^2 Y_{ap}^2 \left(\frac{Y_{ap}+\sqrt{Y_{ap}(Y_{ap}-2M)}}{Y_{ap}-2M+\sqrt{Y_{ap}(Y_{ap}-2M)}}\right)^2 = \\&&
\frac{4a^2b^2}{\left[Y_{ap}-M+\sqrt{Y_{ap}(Y_{ap}-2M)}\right]^2}.
\label{MasterHorizonEq}
\end{eqnarray}

It may be useful to represent the apparent-horizon equation as a sextic of $\mu$, where $s=\sqrt{r^2+b^2}$, $c\equiv\frac{M}{2a}$ and $\mu = \frac{c}{s}$.
\begin{equation}
\mu^2(1-\mu)^2 - \frac{4a^2b^2}{M^2}\mu^4(1-\mu)^2 - \frac{H^2M^2}{4}(1+\mu)^6 = 0.
\label{AHmu}
\end{equation}

While solving Eq. (\ref{MasterHorizonEq}) or Eq.(\ref{AHmu}) by treating them as a polynomial of $Y_{ap}$ or $\mu$ is non-trivial, some important limits can immediately be recovered and discussed. For instance, a static limit, i.e., $H \rightarrow 0$, simply leads to $r_{ap} = 0$. Thus, in the static limit, the apparent horizon coincides with the bounce throat, as expected for a regular black-bounce geometry. In the cosmological limit $M \rightarrow 0$
the central mass is negligible and as a result, $\mu \rightarrow 0$ and $Y = a s$. Therefore, Eq. (\ref{HorizonEq1}) reduces to
\begin{equation}
1-H^2 Y_{ap}^2 = \frac{a^2b^2}{Y_{ap}^2}.
\end{equation}
The above formula suggests that the regularization parameter introduces a finite correction to the usual FLRW cosmological horizon at the order $b^2$, which vanishes smoothly as $b\rightarrow 0$.   

\section{Circular Geodesics, Photon Sphere, Lyapunov exponent and ISCO}\label{geodesic}

The regularized McVittie metric describes a compact object embedded in an expanding FLRW universe. Thus, the spacetime is dynamic and cannot admit a globally conserved energy for geodesic motion. In extension, it can not accommodate strictly circular orbits. However, for most of the astrophysical compact objects, the orbital period is many orders of magnitude smaller than the cosmological expansion time, i.e., $t_{\rm orb}\ll H^{-1}$. During a single orbital period, the fractional change in the scale factor can be considered negligibly small, so that the cosmological expansion may be regarded as effectively frozen. This allows us to use a quasi-static approximation, in which the scale factor is treated as an effective constant parameter, $a_s$. Under this approximation, the regularized McVittie metric assumes a static form
\begin{equation}
d\bar{s}^2 = -f(r)\,dt^2 + g(r)\,dr^2 + Y^2(r)\,d\Omega^2,
\label{StaticMetric}
\end{equation}
where
\begin{eqnarray}
&& f(r) = \left(\frac{1-\mu}{1+\mu}\right)^2 ~~,~~ g(r) = a_s^2(1+\mu)^4,\\&&
\mu(r) = \frac{M}{2a_s\sqrt{r^2+b^2}} ~~,~~ Y(r) = a_s(1+\mu)^2 \sqrt{r^2+b^2}.\label{Yeqn}\nonumber\\
\end{eqnarray}

The metric under a quasi-static approximation admits the Killing vectors $\xi_{(t)}=\partial_t$ and $\xi_{(\phi)}=\partial_\phi$, which yield conserved energy and angular momentum along geodesics. However, it should be emphasized that a quasi-static approximation does not
replace the underlying dynamical spacetime by a genuinely static one. Rather, it describes geodesic motion on a sequence of instantaneous
static hypersurfaces labelled by the slowly varying cosmological scale factor $a_s$. The motion of freely falling test particles is described by the geodesic action

\begin{equation}
S = \frac{1}{2} \int g_{\mu\nu} \frac{dx^\mu}{d\lambda} \frac{dx^\nu}{d\lambda} d\lambda,
\label{GeodesicAction}
\end{equation}

where $\lambda$ denotes an affine parameter along the particle trajectory. The corresponding Lagrangian is $2\mathcal L = g_{\mu\nu}\dot x^\mu\dot x^\nu$, where an overdot denotes differentiation with respect to $\lambda$. We choose to work in an equatorial plane, choosing
$\theta = \frac{\pi}{2}$ and $\dot\theta=0$. Using the quasi-static metric Eq. (\ref{StaticMetric}), the Lagrangian becomes

\begin{equation}
2\mathcal L = -f(r)\dot t^{2} + g(r)\dot r^{2} + Y^2(r)\dot\phi^{2}.
\label{ReducedLagrangian}
\end{equation}

The Euler-Lagrange equations for the above Lagrangian lead to the conserved canonical momenta for the cyclic coordinates

\begin{eqnarray}
&&\frac{d}{d\lambda}\left(f(r)\dot t\right) = 0 ~~ \rightarrow ~~ f(r)\dot t = E, \\&&
\frac{d}{d\lambda}\left(Y^2(r)\dot\phi \right) = 0 ~~ \rightarrow ~~ Y^2(r)\dot\phi = L.
\end{eqnarray}

The radial part of the Euler-Lagrange equations governs the radial motion of massive particles/photons, which is derived as

\begin{equation}\label{RadialGeodesic}
g(r)\ddot r + \frac{1}{2}g'(r)\dot r^2 + \frac{1}{2}f'(r)\dot t^2 - Y(r)Y'(r)\dot\phi^2 = 0.
\end{equation}

We also analyze the normalization of the four-velocity,

\begin{equation}\label{Normalization0}
g_{\mu\nu}\dot x^\mu \dot x^\nu = -\epsilon ~~;~~ \epsilon =
\left\{
\begin{array}{ll}
1,
&
\text{timelike geodesics},
\\
0,
&
\text{null geodesics}.
\end{array}
\right.
\end{equation}

Substituting the metric coefficients in Eq. (\ref{Normalization0}), using the derived conserved quantities $E$ and $L$, we find the familiar set of equations associated with radial motion 
\begin{eqnarray}\label{EffectivePotential0}
&& \dot r^2 + V_{\rm eff}(r) = \frac{E^2}{f(r)g(r)}, \\&&
V_{\rm eff}(r) = \frac{1}{g(r)}\left(\epsilon + \frac{L^2}{Y^2(r)} \right).
\label{EffectivePotential}
\end{eqnarray}                  

The trajectory of a massive particle or a photon is found by solving the first integral for an appropriate effective potential.

\subsection{Massless Particles: Circular Geodesics and Photon Sphere}
We first consider the motion of massless particles, corresponding to $\epsilon = 0$. These are expected to follow circular null geodesics and define the photon sphere, which is the boundary between escaping and captured light rays. It characterizes the optical properties of the geometry. For null geodesics, Eqs. (\ref{EffectivePotential0}) and (\ref{EffectivePotential}) reduce to

\begin{equation}
g(r)\dot r^2 = \frac{E^2}{f(r)} - \frac{L^2}{Y^2(r)}.
\label{NullRadial}
\end{equation}

We introduce an impact parameter $b_{\rm imp}\equiv\frac{L}{E}$, and rewrite Eq.~(\ref{NullRadial}) as

\begin{equation}
\dot r^2 = \frac{E^2}{f(r)g(r)} \left[1 - \frac{Y^2(r)}{f(r)} \frac{f(r)}{Y^2(r)} \right].
\label{NullRadial2}
\end{equation}

A circular null orbit is characterized by a constant radial coordinate, i.e., $\dot r = 0$ and $\ddot r = 0$. The first condition applied to Eq.~(\ref{NullRadial2}) immediately gives

\begin{equation}
b_{\rm imp}^2 = \frac{Y^2(r)}{f(r)}.
\label{ImpactParameter}
\end{equation}

Since the conserved energy $E$ and angular momentum $L$ remain constant along the geodesic, the impact parameter is also constant. Differentiating Eq.~(\ref{ImpactParameter}) with respect to the radial coordinate gives the condition determining circular null geodesics
\begin{equation}
\frac{d}{dr}\left(\frac{f(r)}{Y^2(r)} \right) = \frac{d}{dr}\left[\frac{(1-\mu)^2}{(1+\mu)^6(r^2+b^2)}\right] = 0.
\label{PhotonSphereMcVittie}
\end{equation}

In order to solve for the location of a photon-sphere we rewrite the above equation in terms of $s \equiv \sqrt{r^2+b^2}$, $c\equiv\frac{M}{2a_s}$ and $\mu=\frac{c}{s}$ and find

\begin{equation}
\frac{r}{s} \frac{d}{ds} \left[\frac{s^2(s-c)^2}{(s+c)^6}\right] = \frac{-2r(s-c)\left(s^2-4cs+c^2\right)}{(s+c)^7} = 0.
\label{PhotonSphereFactorized}
\end{equation}

Since $s\geq b >0$ and $c > 0$, the denominator in Eq.~(\ref{PhotonSphereFactorized}) can not vanish. All stationary points are determined by

\begin{equation}
r = 0,~~ s = c,~~ s^2 - 4cs + c^2 = 0.
\end{equation}

The third of these conditions gives

\begin{equation}
s = c(2\pm\sqrt{3}) \Rightarrow r_{\rm ph}^{(\pm)} = \pm \sqrt{\frac{M^2}{4a_s^2}(7\pm4\sqrt{3}) - b^2 }.
\label{PhotonSphereRoots}
\end{equation}

$r = 0$ remains a mathematical stationary point due to the reflection symmetry of the regularized radial coordinate. It points out to a circular null orbit at the bounce surface, subject to the corresponding causal and stability conditions. The second root $s = c$ implies $\mu=1$ and results in a vanishing of the lapse function as

\begin{equation}
f(r)= \left(\frac{1-\mu}{1+\mu}\right)^2 = 0.
\end{equation}

Therefore, this root corresponds to a lapse-zero hypersurface rather than to a regular circular null orbit with finite impact parameter. For the exterior region, where $\mu < 1$ or equivalently $s > c$, only the root $s = c(2+\sqrt{3})$ fall within our scope of interest. The corresponding radial coordinate can be derived as

\begin{equation}
r_{\rm ph} = \sqrt{\frac{M^2}{4a_s^2}(7+4\sqrt{3})-b^2}.
\label{ExteriorPhotonSphere}
\end{equation}

Therefore, the geometry has a photon sphere provided $b \leq \frac{M}{2a_s}(2+\sqrt{3})$. Taking double derivative of $\dot{r}^2$ from Eq. (\ref{NullRadial2}) with respect $r$ we find $8 L^2 a_s^2r_{\rm ph}/81 M^6 (7+4\sqrt{3})$, which is always positive. Thus the photon sphere is unstable. The corresponding physical (or areal) radius of the photon sphere is obtained by substituting Eq.~(\ref{ExteriorPhotonSphere}) into Eq.~(\ref{Yeqn}), yielding

\begin{equation}
Y_{\rm ph}
= a_s \left[1+\mu(r_{\rm ph})\right]^2
\sqrt{r_{\rm ph}^2+b^2}.
\end{equation}

Using the explicit expression for $r_{\rm ph}$, we obtain

\begin{equation}
Y_{\rm ph}
= \left(3-\sqrt{3}\right)^2
\frac{M}{2}\left(2+\sqrt{3}\right)
= 3M.
\end{equation}

Thus, the areal radius of the photon sphere is always equal to $3M$, irrespective of the choice of the time-slicing parameter $a_s$.

\subsection{Lyapunov exponent}

A Lyapunov exponent characterizes the instability of small perturbations around a radial null trajectory in the spacetime metric of a compact object. To determine the Lyapunov exponent, we consider a small radial perturbation, $\delta r$, around the photon orbit at $r=r_{\rm ph}$. We investigate how the effective potential $V_0$, defined by

\begin{equation}
\dot r^2 = \frac{E^2}{f(r)g(r)} - \frac{1}{g(r)}\left(\frac{L^2}{Y^2(r)}\right) = V_0,
\label{RadialFunctionEq1}
\end{equation}
responds to this perturbation. Expanding the radial equation around the photon orbit and retaining the terms linear in $\delta r$, we find the evolution of the perturbation in the form
\begin{equation}
\delta r \sim e^{\Lambda t},
\end{equation}
where $\Lambda$ denotes the Lyapunov exponent characterizing the instability of the photon orbit. This exponent quantifies the rate at which small perturbations grow or decay with time, thereby characterizing the stability of the orbit. The generic expression for the Lyapunov exponent is given by

\begin{equation}
\Lambda= \sqrt{\frac{V_0''(r_{\rm ph})}{2 \dot{t}^2(r_{\rm ph})}}~,
\end{equation}
where the prime denotes a derivative with respect to the radial coordinate $r$. For our case, the Lyapunov exponent is given by,
\begin{eqnarray}
\Lambda&=&\frac{2 a_{s}\sqrt{\frac{M^2}{4a_s^2}(7+4\sqrt{3})-b^2}}{3 \sqrt{3}M^2(2+\sqrt{3})}\nonumber\\
&=&\frac{2 a_{s}r_{\rm ph}}{3 \sqrt{3}M^2(2+\sqrt{3})}.
\end{eqnarray}
For consistency, we consider the limit $a_s\to1$ and $b\to0$. In this limit, the photon-sphere radius and the corresponding Lyapunov exponent reduce to
\begin{eqnarray}
r_{\rm ph}&=&\frac{M}{2}(2+\sqrt{3}),\nonumber\\
\Lambda&=&\frac{2}{3\sqrt{3}M^2(2+\sqrt{3})}
\frac{M}{2}(2+\sqrt{3}),\nonumber\\
&=&\frac{1}{3\sqrt{3}M}.
\end{eqnarray}
Thus, we recover the well-known Lyapunov exponent for the photon sphere of a Schwarzschild black hole.

\subsection{Timelike Circular Geodesics}

We now consider the motion of massive test particles, corresponding to $\epsilon = 1$. The radial equation of motion obtained from Eqs. (\ref{EffectivePotential0}) and (\ref{EffectivePotential}) is

\begin{equation}
g(r)\dot r^2 = \frac{E^2}{f(r)} - \frac{L^2}{Y^2(r)} - 1 \equiv \mathcal{R}(r).
\label{TimelikeRadial}
\end{equation}

A circular timelike orbit at $r=r_c$ is characterized by the absence of radial motion and radial acceleration. Equivalently, the radial function must satisfy
\begin{equation}
\mathcal{R}(r_c)=0,~~ \mathcal{R}'(r_c)=0.
\label{CircularConditions}
\end{equation}

The first condition gives

\begin{equation}
\frac{E^2}{f} = 1+\frac{L^2}{Y^2},
\label{CircularEq1}
\end{equation}

where all quantities are evaluated at $r=r_c$ after imposing the circular-orbit conditions. The second condition gives

\begin{equation}
-\frac{E^2f'}{f^2} + \frac{2L^2Y'}{Y^3} = 0.
\label{CircularEq2}
\end{equation}

Eqs. (\ref{CircularEq1}) and (\ref{CircularEq2}) constitute two algebraic relations for the conserved quantities $E$ and $L$. Following the standard method of elimination and substitution, we derive the conserved quantities as

\begin{equation}
L_c^2 = \frac{Y^3f'}{2fY'-Yf'},~~ E_c^2 = \frac{2f^2Y'}{2fY'-Yf'}.
\label{Lsq}
\end{equation}

For the regularized McVittie metric, with $s(r) = \sqrt{r^2+b^2}$ and $\mu(r)=\frac{M}{2a_s s(r)}$, these can be written as
\begin{equation}
L_c^2 = \frac{a_s M s(1+\mu)^4}{1-4\mu+\mu^2},~~ E_c^2 = \frac{(1-\mu)^4}{(1+\mu)^2\left(1-4\mu+\mu^2\right)},
\end{equation}

For a physically admissible circular orbit, both $E_c^2$ and $L_c^2$ must be positive. In the exterior region, where $0 \leq \mu < 1$, this requirement leads to

\begin{equation}
1-4\mu+\mu^2 > 0 \Rightarrow \mu=2\pm\sqrt{3}.
\label{CircularExistence}
\end{equation}

For $\mu<1$, the relevant boundary is $\mu < 2-\sqrt{3}$. Thus, the family of circular timelike orbits terminates at the radius where $\mu = 2 - \sqrt{3}$.

\subsection{Innermost Stable Circular Orbit}
In order to discuss the stability (under small radial perturbations) of the circular geodesics derived in the previous subsection, we write the first integral Eq. (\ref{TimelikeRadial}) as

\begin{equation}
\dot r^2 = \frac{E^2}{f(r)g(r)} - \frac{1}{g(r)}\left(1+\frac{L^2}{Y^2(r)}\right) = \mathcal{W}(r).
\label{RadialFunctionEq}
\end{equation}

$E$ and $L$ are regarded as fixed constants associated with a given geodesic. A circular orbit at $r=r_c$ corresponds to a double zero of the radial function, i.e., $\mathcal{W}(r_c)=0$ and $\mathcal{W}'(r_c)=0$. The first of these conditions is simply the condition $\dot{r}=0$, while the second ensures that the orbit has no radial acceleration. We now consider a small radial perturbation of the orbit,

\begin{equation}
r(\lambda) = r_c+\delta r(\lambda),~~|\delta r|\ll r_c.
\end{equation}

Expanding the radial function about $r = r_c$ gives

\begin{equation}
\mathcal{W}(r) = \mathcal{W}(r_c) + \mathcal{W}'(r_c)\delta r + \frac{1}{2}\mathcal{W}''(r_c)(\delta r)^2 + \mathcal{O}(\delta r^3).
\end{equation}

Using the circular-orbit conditions, one can derive that for a non-trivial perturbation, the linearized radial equation is

\begin{equation}
\ddot{\delta r} - \frac{1}{2}\mathcal{W}''(r_c) \delta r = \ddot{\delta r} + \omega_r^2\delta r = 0,
\label{LinearRadialPerturbation}
\end{equation}

where $\omega_r^2 \equiv -\frac{1}{2}\mathcal{W}''(r_c)$. A circular orbit is stable when $\mathcal{W}''(r_c) < 0$ and $\omega_r^2 > 0$, for which the radial perturbation undergoes bounded oscillations. On the other hand, $\mathcal{W}''(r_c) > 0$ and $\omega_r^2 < 0$ lead to exponentially growing radial perturbations and therefore signal an unstable circular orbit. The marginally stable orbit is obtained when the radial frequency vanishes, i.e., $\mathcal{W}''(r_c) = 0$, which defines the innermost stable circular orbit (ISCO). For the present geometry, the radial function is defined as in Eq. (\ref{RadialFunctionEq}). Using the circular-orbit values of the conserved quantities derived in
Eq. (\ref{Lsq}), we evaluate the condition $\mathcal{W}''(r_c) = 0$ explicitly as

\begin{equation}
2f'^2Y' - ff''Y' - 3ff' \frac{Y'^2}{Y} + ff'Y'' = 0.
\label{ISCOEq}
\end{equation}

For the regularized McVittie metric under the quasi-static limit, the above equation can be simplified into
\begin{eqnarray}
&& -\frac{4a_s c r^3(s-c)^3\left(s^2-10cs+c^2\right)}{s^6(s+c)^5} = 0,\\&&\nonumber
s\equiv\sqrt{r^2+b^2},~~ c\equiv\frac{M}{2a_s},~~ \mu=\frac{c}{s},~~ \frac{d}{dr} = \frac{r}{s}\frac{d}{ds},\\&&\nonumber
f(r) = \left(\frac{s-c}{s+c}\right)^2,~~ Y(r) = a_s\frac{(s+c)^2}{s}.
\end{eqnarray}
Since $s>0$, $a_s>0$ and $s+c>0$, the ISCO condition reduces to

\begin{equation}
r^3(s-c)^3 \left(s^2-10cs+c^2\right)=0.
\label{ISCOFactorized}
\end{equation}

The condition $s = c$ corresponds to $\mu = 1$, where the lapse function vanishes. This does not represent a regular exterior circular orbit.
The factor $r = 0$ corresponds to the reflection-symmetric bounce surface. The exterior ISCO is determined by the condition

\begin{equation}\label{rootisco}
s^2-10cs+c^2=0 \Rightarrow s_{\rm ISCO}^{(\pm)} = c(5\pm2\sqrt{6}).
\end{equation}

Since the exterior region satisfies $s > c$, the physically relevant branch is

\begin{equation}
s_{\rm ISCO} = \frac{M}{2a_s}(5+2\sqrt{6}).
\label{sISCO}
\end{equation}

Using $s^2=r^2+b^2$, the corresponding coordinate location is

\begin{equation}
r_{\rm ISCO} = \sqrt{\frac{M^2}{4a_s^2}(49+20\sqrt{6})-b^2}.
\label{rISCO}
\end{equation}

The existence of a real exterior ISCO in this branch, therefore, requires

\begin{equation}
b \leq \frac{M}{2a_s}(5+2\sqrt{6}).
\end{equation}

Interestingly, the corresponding areal radius is independent of both the regularization parameter $b$ and the quasi-static scale factor
$a_s$. Indeed, Eq.~(\ref{rootisco}) gives

\begin{equation}
\mu_{\rm ISCO} = \frac{1}{5+2\sqrt6} = 5-2\sqrt6,
\end{equation}

and hence

\begin{equation}\label{YISCO}
Y_{\rm ISCO} = a_s(1+\mu_{\rm ISCO})^2s_{\rm ISCO} = \frac{M}{2} \frac{(6+2\sqrt6)^2}{5+2\sqrt6} = 6M.
\end{equation}

Therefore, a regularization parameter modifies the coordinate location of the exterior ISCO while leaving its areal radius equal to the Schwarzschild value. In the simultaneous limit $b\rightarrow0$ and $a_s\rightarrow1$, Eq. (\ref{rISCO}) reduces to the usual isotropic-coordinate location of the Schwarzschild ISCO.

\section{Regularization of the Cosmological Scale Factor}\label{scalefactor}
The regularization discussed in the preceding sections deals with the central Schwarzschild-like curvature singularity of McVittie geometry, by replacing the isotropic radial coordinate with the regularized function $s(r) = \sqrt{r^2+b^2}$. However, as noted earlier, there remains a second and conceptually distinct source of singularity in the McVittie geometry, namely, the cosmological singularity inherited from the FLRW background at $a(t) \rightarrow 0$. In this section we regularize this as well and introduce a spacetime that is regular both locally and globally. We introduce the regularized scale factor as

\begin{equation}
A(t)=\sqrt{a^{2}(t)+a_{b}^{\,2}},
\label{areg}
\end{equation}

where $a_b > 0$ is a constant regularization scale. The globally regularized McVittie geometry is then written as

\begin{eqnarray}\nonumber
&& d\bar{s}^{2} = - \left( \frac{1-\mu}{1+\mu} \right)^{2} dt^{2} + A^{2}(1+\mu)^{4} \Big[ dr^{2} + s^{2}d\Omega^{2} \Big],\\&&\label{FullyRegularMetric}
\mu(t,r) = \frac{M}{2A(t)s(r)}~,~s(r)=\sqrt{r^{2}+b^{2}}.
\end{eqnarray}

The radius of the extremal two-sphere located at $r = 0$ is now defined as

\begin{equation}
Y_{\rm th} = A(t)b \left( 1+\frac{M}{2A(t)b} \right)^2.
\label{ThroatRadiusFull}
\end{equation}

The geometry consists of two asymptotically cosmological regions joined through a finite-area throat. Rewriting the metric in terms of the coordinate $s$ gives

\begin{equation}\label{morristhorneglobal}
d\bar{s}^2 = - \left( \frac{1-\mu}{1+\mu} \right)^2dt^2 + A^2 (1+\mu)^4 \left[\frac{ds^2}{1-b^2/s^2} + s^2 d\Omega^2 \right],
\end{equation}

which exhibits a radial structure similar to a Morris-Thorne wormhole, through the factor $(1-b^2/s^2)^{-1}$. The quantity $b_{\rm sh}(s) = \frac{b^2}{s}$ can be interpreted as the effective shape function of the wormhole geometry, satisfying the throat condition $b_{\rm sh}(b) = b$. However, unlike the original static Morris-Thorne construction, the present geometry is dynamical, and the throat evolves with the background cosmic expansion while preserving its finite-area character. Therefore, a globally regularized McVittie geometry can be treated as a dynamical generalization of Morris-Thorne wormhole embedded within an expanding universe.    \\

It is straightforward to follow the arguments established in Section \ref{geometry}, differentiate Eq. (\ref{ThroatRadiusFull}) with respect to the regularization parameter and find that the throat radius gives a unique global minimum at $2A(t)b = M$. The minima separate the traversable wormhole and black-bounce branches of the geometry. For $2A(t)b > M$, the extremal two-sphere represents the throat of a traversable wormhole connecting the two asymptotically cosmological regions, whereas $2A(t)b < M$ corresponds to the black-bounce branch in which the regular throat is hidden behind trapped surfaces. Therefore, the introduction of a regularized scale factor only modifies the dynamical evolution of the throat but leaves the intrinsic properties of the geometric patch unchanged.   \\

\begin{figure}[h]
	\begin{center}
		\includegraphics[angle=0, width=0.35\textwidth]{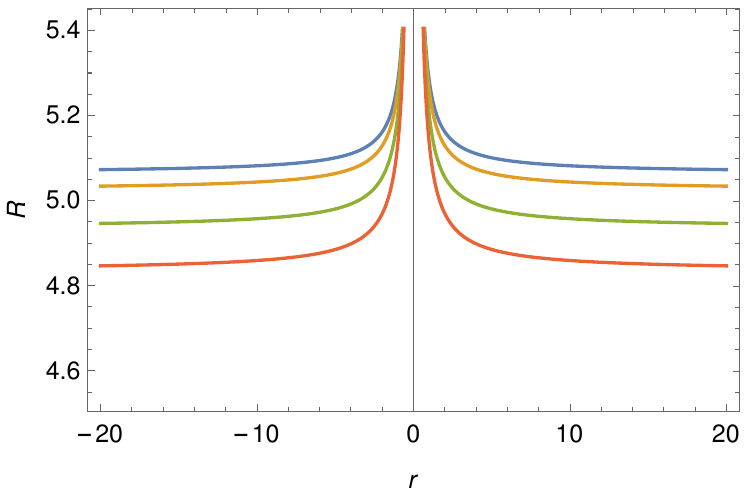}
	\end{center}
	\caption{{\small Ricci scalar as a function of $r$. Curves of different color signify different values of time, used as different snaps of the evolution}}
	\label{Ricci_2}
\end{figure}

The regularity of the geometry can be seen directly from the explicit expression for the Ricci scalar as in Eq. (\ref{ricccccccci2}). Since the regularization scales satisfy $A(t)\geq a_b>0$ and $s(r)=\sqrt{r^2+b^2}\geq b>0$, the quantity $\mu(t,r)=\frac{M}{2A(t)s(r)}$ remains finite throughout the spacetime. We confirm this explicitly by evaluating the Ricci scalar $\mathcal{R}$ for a representative regularized scale factor, $A(t)=\sqrt{a_0^2+a(t)^2}$ with $a(t)\sim e^{H_0t}$, and displaying its dependence on $r$ and $t$ in FIG.~\ref{Ricci_2} (the plot is for the scalar as a function of $r$ while different values of time is used as different snaps of the evolution). The non-singular behavior can also be deduced from the denominator of the Ricci scalar in Eq. (\ref{ricccccccci2}), which does not go to zero for any value of $r$. The Ricci scalar is therefore finite in the limit $r \rightarrow 0$ as well as in the limit $a(t)\rightarrow0$.

\begin{widetext}
\footnotesize
\begin{equation}
\begin{aligned}
\mathcal{R}
={}&\frac{2\Big[4(b^2+r^2)a_b^2-M^2 +4(b^2+r^2)a(t)^2\Big]}{\sqrt{b^2+r^2}\,\big[a_b^2+a(t)^2\big]^{5/2}\Big[M-2\sqrt{b^2+r^2}\sqrt{a_0^2+a(t)^2}\Big]^2} \times \frac{1}{\Big[M+2\sqrt{b^2+r^2}\sqrt{a_0^2+a(t)^2}\Big]^6}\Bigg\{192(b^2+r^2)^3\Big[3M+\sqrt{b^2+r^2}
\\[1mm]
& \sqrt{a_b^2+a(t)^2}\Big]a''(t)a(t)^9 + 64(b^2+r^2) \Bigg[\Big(M-\sqrt{b^2+r^2}\sqrt{a_b^2+a(t)^2}\Big)b^2+3(b^2+r^2)^2 \Big(M+\sqrt{b^2+r^2}\sqrt{a_b^2+a(t)^2}\Big) a'(t)^2 \Bigg]a(t)^8 
\\[1mm]
&
+ 48(b^2+r^2)^2\Bigg[ 16(b^2+r^2) \Big(3M+\sqrt{b^2+r^2}\sqrt{a_b^2+a(t)^2}\Big)a_b^2 +5M^2 \Big(2M+3\sqrt{b^2+r^2}\sqrt{a_b^2+a(t)^2}\Big) \Bigg]a''(t)a(t)^7+16 \Bigg\{ \Big[ -16(b^2 
\\[1mm]
&
+r^2)\Big(\sqrt{b^2+r^2}\sqrt{a_b^2+a(t)^2}-M\Big)a_b^2-3M^2\sqrt{b^2+r^2}\sqrt{a_b^2+a(t)^2}\Big]b^2 + 3(b^2+r^2)^2\Big[ 8a_b^2(b^2+r^2)\Big(3M+2\sqrt{b^2+r^2}\sqrt{a_b^2+a(t)^2}\Big)
\\[1mm]
&
-5M^2 \Big(2M+\sqrt{b^2+r^2}\sqrt{a_b^2+a(t)^2} \Big) \Big] a'(t)^2 \Bigg\}a(t)^6 +36(b^2+r^2) \Bigg[ 32(b^2+r^2)^2 \Big(3M+\sqrt{b^2+r^2} \sqrt{a_b^2+a(t)^2}\Big)a_b^4+20M^2(b^2+r^2)
\\[1mm]
&
\Big(2M+3\sqrt{b^2+r^2}\sqrt{a_b^2+a(t)^2}\Big)a_b^2 + M^4 \Big(M+5\sqrt{b^2+r^2} \sqrt{a_b^2+a(t)^2}\Big) \Bigg]a''(t)a(t)^5 +12\Bigg\{ (b^2+r^2) \Bigg[ 96(b^2+r^2)^2 \Big(2M+\sqrt{b^2+r^2}
\\[1mm]
&
\sqrt{a_b^2+a(t)^2}\Big)a_b^4+20M^2(b^2+r^2) \Big(\sqrt{b^2+r^2}\sqrt{a_b^2+a(t)^2}-2M\Big)a_b^2 -M^4 \Big(7M+25\sqrt{b^2+r^2}\sqrt{a_b^2+a(t)^2}\Big)\Bigg]a'(t)^2-4a_b^2b^2 \Big[8(b^2
\\[1mm]
&
+r^2)\Big(\sqrt{b^2+r^2}\sqrt{a_b^2+a(t)^2}-M\Big)a_b^2 +3M^2\sqrt{b^2+r^2}\sqrt{a_b^2+a(t)^2} \Big] \Bigg\}a(t)^4+3\Bigg[256(b^2+r^2)^3 \Big(3M+\sqrt{b^2+r^2}\sqrt{a_b^2+a(t)^2}\Big)a_b^6
\\[1mm]
&
+240M^2(b^2+r^2)^2 \Big(2M+3\sqrt{b^2+r^2}\sqrt{a_b^2+a(t)^2}\Big)a_b^4 +24M^4(b^2+r^2) \Big(M+5\sqrt{b^2+r^2}\sqrt{a_b^2+a(t)^2}\Big)a_b^2+M^6\sqrt{b^2+r^2}\sqrt{a_b^2+a(t)^2}
\\[1mm]
&
\Bigg]a''(t)a(t)^3 + \Bigg\{3 \Bigg[128(b^2+r^2)^3 \Big(5M+2\sqrt{b^2+r^2}\sqrt{a_b^2+a(t)^2}\Big)a_b^6+80M^2(b^2+r^2)^2 \Big(2M+5\sqrt{b^2+r^2}\sqrt{a_b^2+a(t)^2}\Big)a_b^4 - 8M^4(b^2+r^2) 
\\[1mm]
&
\Big(2M+5\sqrt{b^2+r^2}\sqrt{a_b^2+a(t)^2}\Big)a_b^2 -3M^6 \sqrt{b^2+r^2}\sqrt{a_b^2+a(t)^2} \Bigg]a'(t)^2 -16a_b^4b^2 \Big[ 16(b^2+r^2) \Big(\sqrt{b^2+r^2}\sqrt{a_b^2+a(t)^2}-M\Big)a_b^2 + 9M^2 
\\[2mm]
&
\sqrt{b^2+r^2} \sqrt{a_b^2+a(t)^2}\Big] \Bigg\}a(t)^2+3a_b^2\Bigg[ 64(b^2+r^2)^3 \Big(3M+\sqrt{b^2+r^2}\sqrt{a_b^2+a(t)^2}\Big)a_b^6 + 80M^2(b^2+r^2)^2 \Big(2M+3\sqrt{b^2+r^2}\sqrt{a_b^2+a(t)^2}\Big)
\\[1mm]
&
a_b^4+12M^4(b^2+r^2)\Big(M+5\sqrt{b^2+r^2}\sqrt{a_b^2+a(t)^2}\Big)a_b^2 +M^6\sqrt{b^2+r^2}\sqrt{a_b^2+a(t)^2} \Bigg]a''(t)a(t) + 3 \Bigg[ 64(b^2+r^2)^3 \Big(3M+\sqrt{b^2+r^2}
\\[1mm]
&
\sqrt{a_b^2+a(t)^2}\Big)a_b^8 + 80M^2(b^2+r^2)^2 \Big(2M+3\sqrt{b^2+r^2}\sqrt{a_b^2+a(t)^2}\Big)a_b^6 +12M^4(b^2+r^2) \Big(M+5\sqrt{b^2+r^2}\sqrt{a_b^2+a(t)^2}\Big)a_b^4 
\\[1mm]
&
+ M^6\sqrt{b^2+r^2}\sqrt{a_b^2+a(t)^2}\,a_b^2 \Bigg]a'(t)^2-16a_b^6b^2 \Big[ 4(b^2+r^2) \Big(\sqrt{b^2+r^2}\sqrt{a_b^2+a(t)^2}-M\Big)a_b^2 +3M^2\sqrt{b^2+r^2}\sqrt{a_b^2+a(t)^2} \Big] \Bigg\}.
\end{aligned}
\label{ricccccccci2}
\end{equation}
\end{widetext}

\begin{figure}[!htbp]
	\begin{center}
		\includegraphics[angle=0, width=0.35\textwidth]{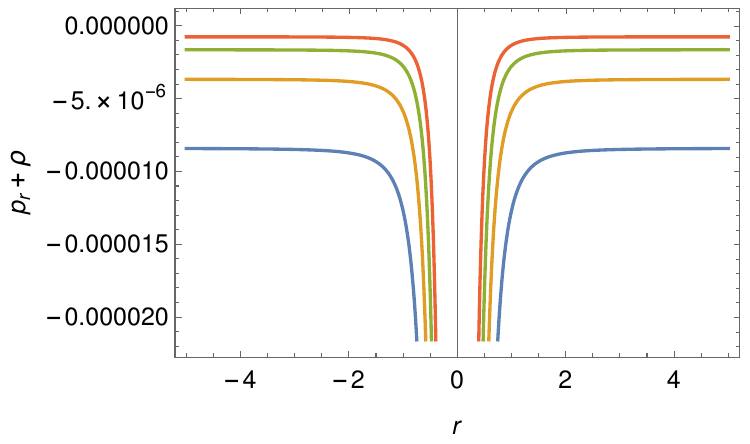}
		\includegraphics[angle=0, width=0.35\textwidth]{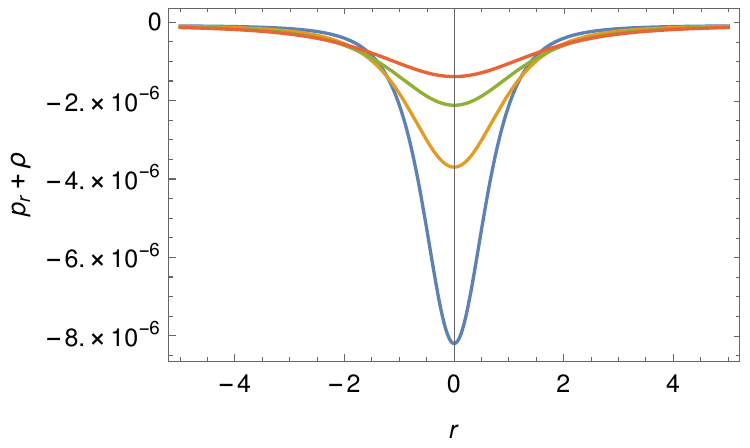}
		\includegraphics[angle=0, width=0.35\textwidth]{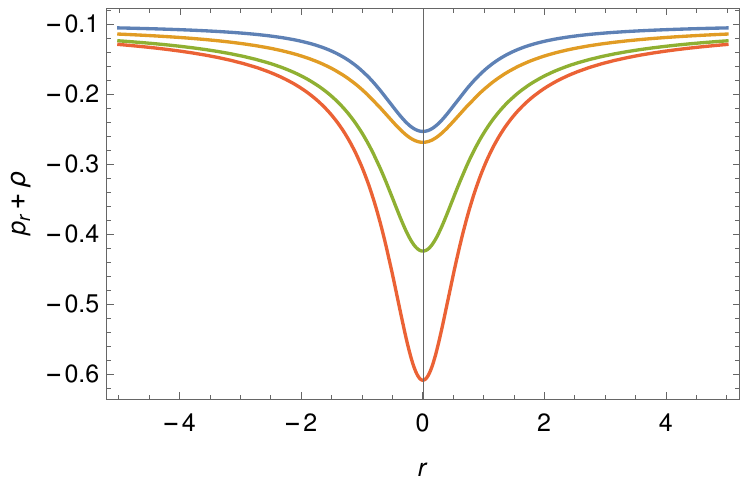}
		\includegraphics[angle=0, width=0.35\textwidth]{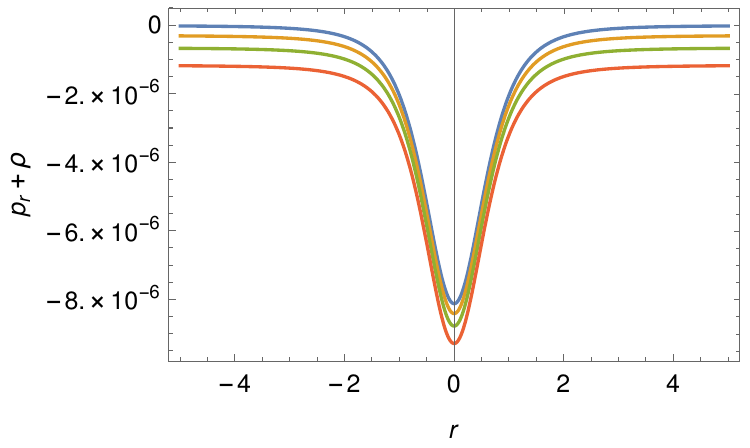}
		\includegraphics[angle=0, width=0.35\textwidth]{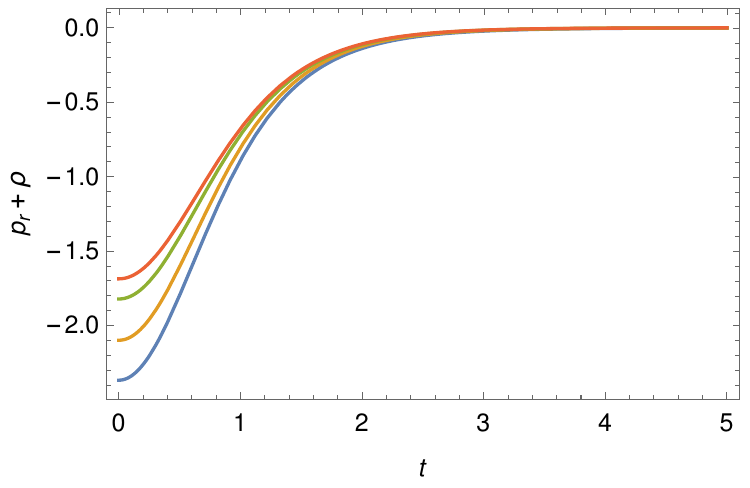}
	\end{center}
	\caption{{\small The profile of left-hand side of NEC, $\rho + p_r$, for four different setups. Graph on top : $\rho + p_r$ as a function of $r$, for different values of coordinate time $t$. Graph second from top : $\rho + p_r$ as a function of $r$, for different values of the regularization parameter $b$. Graph third from top : $\rho + p_r$ as a function of $r$ for different values of the McVittie mass $M$. Graph fourth from top : $\rho + p_r$ as a function of $r$ for different values of the cosmological regularization scale $a_0$. Graph below : $\rho + p_r$ as a function of time, for different shells, i.e., different values of $r$.}}
	\label{NEC_2}
\end{figure}

The regularity of the curvature invariants does not, by itself, imply that the effective matter content satisfies the standard energy conditions. In particular, the NEC provides a useful description of whether or not the matter distribution can sustain the regularized geometry. In the present construction, the NEC behaves in a qualitatively similar manner compared to the radially regularized geometry discussed earlier. We show this in FIG. \ref{NEC_2}, plotting the left-hand side of NEC, i.e., $\rho + p_r$ for five different setups. First, we plot it as a function of $r$, for different values of coordinate time $t$ (graph on top). We also plot the NEC for different values of the regularization parameters $b$ (graph second from the top) and the McVittie mass $M$ (graph third from the top). We add one plot in this case, that of $\rho + p_r$ vs $r$ for different values of the second regularization parameter $a_b$ (graph fourth from the top). Finally, we plot the NEC as a function of time, for different shells, i.e., different values of $r$ (graph below). For all the cases considered, it can be seen that there is a general violation of the Null Energy Condition.  \\

The time-regularized Mcvittie metric naturally interpolates between different geometric patches under appropriate limits of the two regularization parameters. In the limit $a_b \rightarrow 0$, the regularized scale factor reduces to the FLRW scale factor, $A(t) \rightarrow a(t)$, and one recovers the regularized McVittie spacetime developed in the earlier parts of the work. In the limit $b \rightarrow 0$, the radial regularization is eliminated as well, and one has the ordinary McVittie metric. In the static limit $A(t) \rightarrow 1$, there is no cosmological evolution, and the spacetime reduces to a black-bounce geometry. Taking simultaneous limits $b \rightarrow 0$ and $A(t) \rightarrow 1$ gives the Schwarzschild solution. Therefore, this construction provides a unified geometric framework encompassing the Schwarzschild, Simpson-Visser, McVittie, and regularized McVittie spacetimes as different members of a single geometric patch depending on a two-parameter family.   \\

We also show that a regularization of the cosmological scale factor modifies the locations of the apparent horizon. We recall that the location of apparent horizons is derived from 
\begin{eqnarray}\label{AHConditionReg}
&& g^{ab}\nabla_a Y \nabla_b Y = 0~,~ Y(t,r) = A(t)(1+\mu)^2 s, \\&&
\mathcal H = \frac{\dot{A}}{A}~~,~~ \mu = \frac{M}{2A(t)s}.
\end{eqnarray}

Substituting the metric coefficients from Eq. (\ref{FullyRegularMetric}), and taking into account $\dot M = 0$, we derive that
\begin{equation}
\frac{r_{\rm AH}^{\,2}}{r_{\rm AH}^{\,2}+b^2} = \mathcal H^2 Y_{\rm AH}^{\,2} \left(\frac{1+\mu}{1-\mu} \right)^2.
\label{AHNoAccretionReg}
\end{equation}

Expressing the result in terms of the regularized radial coordinate $s=\sqrt{r^2+b^2}$ gives us
\begin{equation}
1- \frac{b^2}{s_{\rm AH}^{\,2}} = \mathcal H^2 Y_{\rm AH}^{\,2} \left(\frac{1+\mu}{1-\mu}\right)^2.
\label{AHsReg}
\end{equation}

We follow along the technical steps discussed in Section $V$ (Eqs. (\ref{quadratic}) and (\ref{sintermsofr})) and express the regularized areal radius as a quadratic of $s$ to find

\begin{eqnarray}
&& As^2+(M-Y)s+\frac{M^2}{4A} = 0, \\&&
s = \frac{Y-M+\sqrt{Y(Y-2M)}}{2A}, \\&&
\mu = \frac{M}{Y-M+\sqrt{Y(Y-2M)}}.
\end{eqnarray}

As a result, the apparent horizon condition becomes

\begin{eqnarray}\nonumber
&& 1 - \mathcal H^2 Y_{\rm AH}^{\,2}\left(\frac{Y_{\rm AH} + \sqrt{Y_{\rm AH}(Y_{\rm AH}-2M)}}{Y_{\rm AH}-2M+\sqrt{Y_{\rm AH}(Y_{\rm AH}-2M)}}\right)^2 = \\&&
\frac{4A^2b^2}{\left[Y_{\rm AH} - M + \sqrt{Y_{\rm AH}(Y_{\rm AH}-2M)}\right]^2}.
\label{MasterHorizonRegularized}
\end{eqnarray}

It is also useful to rewrite the apparent-horizon condition in terms of the regularized radial coordinate $s$, $\mathbf{c} \equiv \frac{M}{2A(t)}$, and $\mu=\frac{\mathbf{c} }{s}$, as
\begin{equation}
(s^2-b^2)(s-\mathbf{c} )^2 - \mathcal{H}^2 A^2(s+\mathbf{c} )^6 = 0.
\label{AHsexticCompact}
\end{equation}
At any fixed cosmological time, the apparent-horizon locations are determined by a sixth-order polynomial equation in $s$. Although the general sextic does not provide a useful closed-form expression for the horizon radii, two useful limiting cases of the horizon equation can be discussed here. First, in an infinitesimally small mass limit, i.e., $M \rightarrow 0$, $\mathbf{c} \rightarrow 0$, Eq.~(\ref{AHsexticCompact}) reduces to
\begin{equation}
(s^2-b^2)s^2 = \mathcal{H}^2A^2s^6.
\end{equation}
For $s \neq 0$, this is a quadratic equation for $s^2$ with solutions
\begin{equation}
s^2_{\rm AH} = \frac{1\pm\sqrt{1-4\mathcal{H}^2A^2b^2}}{2\mathcal{H}^2A^2}.
\end{equation}
It is evident that the existence of real apparent horizons requires $2\mathcal{H}A b\leq1$. Similarly, when the radial regularization parameter is infinitesimally small, i.e, $b \rightarrow 0$, the sextic equation reduces to a cubic equation
\begin{equation}
s^2(s-\mathbf{c} )^2 = \mathcal{H}^2A^2(s+\mathbf{c} )^6.
\end{equation}

The complete regularization provides a simple extension of the radially regularized McVittie geometry. The formal correspondence with the radially regularized case can be expressed through $a(t) \rightarrow A(t)$ and $H(t)\rightarrow\mathcal{H}(t)$, while the radial regularization continues to be governed independently by the scale $b$. The two regularization scales play distinct roles: $b$ replaces the singular central core by a finite-area bounce surface, whereas $a_b$ prevents the cosmological scale factor from reaching zero. This distinction is also reflected in the apparent-horizon structure. Since $A(t) \geq a_b > 0$, the effective expansion rate $\mathcal{H}(t)$ remains finite for a regular choice of the underlying scale factor, and the horizon equation evolves without encountering the singular $a(t)=0$ limit. At the same time, the radial horizon structure continues to depend on the compact-object parameters and the bounce scale through the sextic relation derived above. The resulting spacetime retains the essential McVittie nature : a compact object embedded in an evolving cosmological background, while avoiding both of its singularities by finite geometric scales.

\section{Conclusion}
\label{Conclusion}

In this article, a regularized McVittie spacetime has been constructed in which the two singularities usually associated with the compact-object core and the cosmological background are treated by introducing finite geometric scales. The radial regularization $s(r)=\sqrt{r^2+b^2}$ removes the singular center and replaces it with a finite-area extremal two-sphere, while the redefinition $a(t)\rightarrow A(t)=\sqrt{a^2(t)+a_b^2}$ prevents the cosmological scale factor from vanishing. The two parameters therefore provide independent control over the local and cosmological sectors of the geometry.  \\

This provides a unified geometric picture in which a cosmologically embedded compact object can exhibit black-hole, black-bounce, or wormhole-like behaviour, depending on the relative scales of the mass and the radial regularization. In particular, the condition $2A(t)b=M$ separates the throat and black-bounce regimes. The finite areal radius of the extremal two-sphere and the corresponding null-congruence show that the regular core is not just a coordinate replacement but represents a modification of the central geometry. It has been discussed that the regularized geometry can effectively be interpreted as a time-evolving analogue of Morris-Thorne wormhole embedded in an evolving cosmological background.   \\

The regularity is achieved at the price of an \textit{exotic} effective stress-energy tensor. It is proven that the geometry can only be supported by an imperfect, anisotropic fluid which exhibits a clear violation of the Null energy condition ; typically associated with non-singular throat or bounce geometries. The causal structure is further analyzed by deriving the apparent-horizon equation. Solutions to the apparent horizon equation are usually found implicitly, through higher-order polynomial equations, and they depend on the compact object
mass, the radial regularization scale, and the cosmological expansion. The horizon structure is therefore found to evolve with the background rather than being determined solely by the local compact-object geometry. \\

Irrespective of these departures from a standard McVittie metric, the regularized geometry is found to preserve a remarkable degree of the Schwarzschild core character. This is discussed at length by deriving the photon spheres and marginally stable circular orbits in a quasi-static limit. While the photon spheres and ISCO are modified in their isotropic-coordinate locations, their corresponding areal radii retain the
values associated with the Schwarzschild geometry. The regularization, therefore, affects the coordinate representation alone without necessarily shifting the intrinsic areal scales. The cosmological regularization completes this construction by allowing the regular compact-object geometry to be embedded in a background which itself remains finite at all times.  \\

A useful consequence of the radial regularization is that the restrictive matter structure of standard McVittie spacetime is naturally resolved. In a standard McVittie geometry, the no-accretion condition leads to a homogeneous energy density, while the pressure remains inhomogeneous, $p=p(t,r)$, resulting in a rather non-trivial issue related to the equation of state of the constituent energy-momentum tensor. In the present case, the regularization introduces an inhomogeneity in the effective energy density. Importantly, this modification is localized around the regularized core: the anisotropic contribution becomes negligible at large radius, and the ordinary McVittie behaviour is recovered asymptotically; likewise, the limit $b\rightarrow0$ restores the standard McVittie form.  \\

Several questions remain open. In particular, a systematic exploration of the full parameter space would be required to classify the possible
horizon and geodesic configurations and to quantify the associated energy-condition violations. A fully dynamical treatment of particle
and photon motion beyond the quasi-static approximation would also be desirable. Such an analysis could clarify whether the regular core and
the cosmological regularization leave observable imprints in lensing, photon-ring or shadow formation, or in the propagation of matter and
radiation. The present construction nevertheless provides a simple framework in which the local regularization of a compact object and the
global regularization of its cosmological environment is treated as part of the same spacetime geometry. Furthermore, the perturbative stability of the regularized spacetime and its response to scalar, electromagnetic, and gravitational perturbations remain important directions for future work. Such analyses could clarify whether the regularization of both the central compact object and the cosmological background leads to distinctive observable signatures. The present construction nevertheless provides a unified framework in which the local regularization of a compact object and the global regularization of its cosmological environment are incorporated within a single, self-consistent spacetime geometry.

\section*{Acknowledgments}

 S.C. acknowledges the IUCAA for providing the facility and support under the visiting associateship program. Acknowledgment is also given to the Vellore Institute of Technology for the financial support through its Seed Grant (No. SG20230027), 2023. C.S. thanks the local hospitality
at SINP Kolkata,
where a part of this work has been done.


\begin{thebibliography}{99}

\bibitem{planck1} Planck Collaboration, N. Aghanim et. al., Astron. Astroph. {\bf 641}, A6 (2020).

\bibitem{planck2} P. A. R. Ade et al. Planck 2015 results, Astron. Astrophys. {\bf 594}, A13 (2016).

\bibitem{obs1} D. J. Eisenstein et al. Astrophysical Journal, {\bf 633} : 560, (2005).

\bibitem{obs2} L. Anderson et al. Mon. Not. Roy. Astron. Soc., {\bf 441} : 24 (2014).

\bibitem{obs3} D. Brout et al. Astrophysical Journal, {\bf 938} : 110 (2022).

\bibitem{tension1} A. R. Liddle, Phys. Rep. {\bf 307}, 53 (1998).

\bibitem{tension2} K. Nakayama, F. Takahashi and T. T. Yanagida, Phys. Lett. B {\bf 725}, 111 (2013).

\bibitem{tension3} S. Tsujikawa, Class. Quant. Grav. {\bf 30}, 214003 (2013).

\bibitem{tension4} E. Di Valentino et. al., Class. Quantum Grav. {\bf 38}, 153001 (2021).

\bibitem{tension5} E. Di Valentino et. al., Phys. Dark Univ. {\bf 49}, 101965 (2025).

\bibitem{tension6} R-G Cai and S-J Wang, Res. Astron. Astrop. {\bf 26}, 084011 (2026).

\bibitem{bounce1} M. Novello and S. E. Perez Bergliaffa, Phys. Rept. {\bf 463}(4), 127 (2008).

\bibitem{bounce2} D. Battefeld and P. Peter, Phys. Rept. {\bf 571}, 1 (2015).

\bibitem{bounce3} S. D. Odintsov and V. K. Oikonomou, Phys. Rev. D {\bf 92}, 024016 (2015).

\bibitem{bounce4} M. Koehn, J-L Lehners and B. Ovrut, Phys. Rev. D {\bf 93}, 103501 (2016).

\bibitem{bounce5} J. D. Barrow and C. Ganguly, Int. J. Mod. Phys. D {\bf 26}, 1743016 (2017).

\bibitem{bounce6} V. P. Frolov and A. Zelnikov, Phys. Rev. D {\bf 104}, 104060 (2021).

\bibitem{bounce7} K. A. Meissner and R. Penrose, arXiv : 2503.24263v1 [gr-qc].

\bibitem{bhbounce1} B. J. Carr and A. A. Coley, Int. J. Mod. Phys. D {\bf 20}, 2733 (2011).

\bibitem{bhbounce2} T. Clifton, B. J. Carr and A. A. Coley, Class. Quant. Grav. {\bf 34}, 135005 (2017).

\bibitem{bhbounce3} B. Carr and J. Silk, Mon. Not. R. Astron. Soc. {\bf 478}, 3756 (2018).

\bibitem{bhbounce4} B. Carr and F. Kuhnel, Ann. Rev. Nucl. Part. Sc. {\bf 70}, annurev (2020), arXiv:2006.02838 [astro-ph.CO].

\bibitem{bhbounce5} A. M. Green and B. J. Kavanagh, Jour. Phys. G. Nucl. Phys. {\bf 48}, 043001 (2021).

\bibitem{bhbounce6} K. Jedamzik, J. Cosm. Astropart. Phys. {\bf 2020}, 022 (2020).

\bibitem{mcvittie} G. C. McVittie, Mon. Not. Roy. Astron. Soc. {\bf 93}, 325 (1933).

\bibitem{mcv1} P. Kustaanheimo and B. Qvist, Commentationes Physico-Mathematicae {\bf 13}, 1 (1948), reprinted in Gen. Relativ. Gravit. {\bf 30}(4), 663 (1998).

\bibitem{mcv2} B. C. Nolan, Phys. Rev. D {\bf 58}, 064006 (1998); Class. Quant. Grav. {\bf 16}, 1227 Class. Quant. Grav. 16, 3183 (1999).

\bibitem{mcv3} K. Bolejko, M. N. Celerier and A. Krasinski, Class. Quant. Grav. {\bf 28}, 164002 (2011).

\bibitem{mcv4} N. Kaloper, M. Kleban and D. Martin, Phys. Rev. D {\bf 81}, 104044 (2010).

\bibitem{mcv5} M. Carrera and D. Giulini, Phys. Rev. D {\bf 81}, 043521 (2010).

\bibitem{mcv6} K. Lake and M. Abdelqader, Phys. Rev. D {\bf 84}, 044045 (2011).

\bibitem{mcv7} C. J. Gao, X. Chen, V. Faraoni and Y. G. Shen, Phys. Rev. D {\bf 78}, 024008 (2008).

\bibitem{mcv8} V. Faraoni, C. Gao, X. Chen and Y. G. Shen, Phys. Lett. B {\bf 671}, 7 (2009).

\bibitem{nosing1} J. M. Bardeen, Proceedings of International Conference GR5, 1968, Tbilisi, USSR, p. 174.

\bibitem{nosing2} T. A. Roman and P. G. Bergmann, Phys. Rev. D {\bf 28}, 1265 (1983).

\bibitem{nosing3} S. A. Hayward, Phys. Rev. Lett. {\bf 96}, 031103 (2006).

\bibitem{nosing4} V. P. Frolov, JHEP {\bf 1405}, 049 (2014).

\bibitem{nosing5} V. P. Frolov, Phys. Rev. D {\bf 94}(10), 104056 (2016).

\bibitem{nosing6} V. P. Frolov and A. Zelnikov, Phys. Rev. D {\bf 95}(12), 124028 (2017).

\bibitem{nosing7} R. Carballo-Rubio, F. Di Filippo, S. Liberati, C. Pacilio and M. Visser, JHEP {\bf 1807}, 023 (2018).

\bibitem{nosing8} R. Carballo-Rubio, F. Di Filippo, S. Liberati and M. Visser, Phys. Rev. D {\bf 98}, 124009 (2018).

\bibitem{nosing9} A. De Felice and S. Tsujikawa, Phys. Rev. Lett. {\bf 134}, 081401 (2025).

\bibitem{penrose} R. Penrose, Phys. Rev. Lett. {\bf 14}, 57 (1965) ; Nuovo Cimento Rivista Serie {\bf 1} (1969).

\bibitem{fR} A. M. Nzioki, S. Carloni, R. Goswami, and P. K. S. Dunsby, Phys. Rev. D {\bf 81}, 084028 (2010); J. A. R. Cembranos, A. Cruz-Dombriz and B. Montes-Nunez, JCAP {\bf 1204}, 021 (2012) ; S. Capozziello, M. De Laurentis and A. Stabile, Class. Quant. Grav. {\bf 27}, 165008 (2010).

\bibitem{wh1} M. S. Morris and K. S. Thorne, Am. J. Phys. {\bf 56}, 395 (1988).

\bibitem{wh2} M. S. Morris, K. S. Thorne and U. Yurtsever, Phys. Rev. Lett. {\bf 61}, 1446 (1988).

\bibitem{wh3} M. Visser, Phys. Rev. D {\bf 39}, 3182 (1989).

\bibitem{wh4} M. Visser, Nucl. Phys. B {\bf 328}, 203 (1989).

\bibitem{wh5} M. Visser, S. Kar and N. Dadhich, Phys. Rev. Lett. {\bf 90}, 201102 (2003).

\bibitem{wh6} D. Hochberg and M. Visser, Phys. Rev. D {\bf 56}, 4745 (1997).

\bibitem{wh7} E. Poisson and M. Visser, Phys. Rev. D {\bf 52}, 7318 (1995).

\bibitem{wh8} D. Hochberg and M. Visser, Phys. Rev. Lett. {\bf 81}, 746 (1998).

\bibitem{wh9} J. G. Cramer, R. L. Forward, M. S. Morris, M. Visser, G. Benford and G. A. Landis, Phys. Rev. D {\bf 51}, 3117 (1995).

\bibitem{simpsonvisser} A. Simpson and M. Visser, JCAP {\bf 1902}, 042 (2019).

\bibitem{scsk} S. Chakrabarti and S. Kar, Phys. Rev. D {\bf 104}, 024071 (2021).

\bibitem{faraonimcv} V. Faraoni and A. Jacques, Phys. Rev. D {\bf 76}, 063510 (2007).

\bibitem{harada} T. Harada, H. Maeda and T. Sato, Phys. Lett. B {\bf 833}, 137332 (2022).

\bibitem{abdalla} E. Abdalla, N. Afshordi, M. Fontanini, D. C. Guariento and E. Papantonopoulos, Phys. Rev. D {\bf 89}, 104018 (2014).

\bibitem{afshordi} N. Afshordi1, M. Fontanini and D. C. Guariento, Phys. Rev. D {\bf 90}, 084012 (2014).

\bibitem{nolan} B. C Nolan, Class. Quant. Grav. {\bf 42}, 235019 (2025).

\bibitem{necmcvittie} M. Cadoni, L. de Lima, M. Pitzalis, D. C. Rodrigues and A. P. Sanna, Phys. Rev. D {\bf 113}, 084049 (2026).

\bibitem{rc} A. Raychaudhuri, Phys. Rev. {\bf 98}, 1123 (1955).

\bibitem{ncc} C. W. Misner and J. A. Wheeler, Ann. Phys. (N.Y.) {\bf 2}, 525 (1957).

\bibitem{nec1} C. A. Kolassis, N. O. Santos, and D. Tsoubelis, Class. Quant. Grav. {\bf 5}, 1329 (1988).

\bibitem{nec2} O. M. Pimentel, F. D. Lora-Clavijo and G. A. Gonzalez, Gen. Relativ. Gravit. {\bf 48}, 124 (2016).





\end{thebibliography}
\end{document}